\documentclass[article]{aa} 
\usepackage[varg]{txfonts}
\usepackage{graphicx}
\usepackage{ulem}
\usepackage{txfonts}
\usepackage{natbib}
\usepackage{multirow}
\usepackage[usenames]{color}

\usepackage{enumerate}
\usepackage{booktabs}

\usepackage[english]{babel}

\newlength{\benz}
\newcommand{\mic}{\rm{\,\mu m}}

\newcommand{\ch}[2]{$\rm{C}_{#1}\rm{H}_{#2}$}
\newcommand{\chp}[3]{${\rm{C}_{#1}\rm{H}_{#2}}^{#3}$}

\newcommand{\hh}{$\rm{H}_2$~}
\newcommand{\hhp}{$\rm{H}_2$} 

\newcommand{\nH}{{n_{\rm H}}}

\newcommand{\Av}{{$A_{\rm V}$}}

\newcommand{\eIR}{e_{\rm{IR}}}
\newcommand{\kIR}{k_{\rm{IR}}}
\newcommand{\kdiss}{k_{\rm{diss}}}

\begin{document}

   \title{Evolution of PAHs in photodissociation regions}
   \subtitle{Hydrogenation and charge states}
   \author{J. Montillaud\inst{1,2,3}
          \and
          C. Joblin\inst{1,2}
                   \and 
          D. Toublanc\inst{1,2}
          }
 
   \institute{
   Universit\'e de Toulouse, UPS-OMP, IRAP, Toulouse, France
   \and
   CNRS, IRAP, 9 Av. colonel Roche, BP 44346, 31028, Toulouse Cedex 4, France
   \and
   Department of Physics, P.O.Box 64, FI-00014, University of Helsinki, Finland
   }

\date{Received ????; Accepted????}
\offprints{C.~Joblin, \email{christine.joblin@irap.omp.eu}} 
   \abstract
   {Various studies have emphasized variations of the charge state and composition of the interstellar polycyclic aromatic hydrocarbon
   (PAH) population in  photodissociation regions (PDRs). These changes are expected to impact the energetics and chemistry in
   these regions calling for a quantitative description.
   }
   {We aim to model the spatial evolution of the charge and hydrogenation states of PAHs in PDRs. We focus on the 
   specific case of the north-west (NW) PDR of NGC 7023, for which many observational constraints are available. We also 
   discuss the case of the diffuse interstellar medium (ISM).
   }
   {The physical conditions in NGC 7023 NW are modelled using a state-of-the-art PDR code. We then use a new PAH chemical 
   evolution model that includes recent experimental data on PAHs and describes multiphoton events. We consider a family 
   of compact PAHs bearing up to 96 carbon atoms.
   }
   {The calculated ionization ratio is in good agreement with the observed ratio in NGC 7023 NW. Within the PDR, PAHs 
   evolve into three major populations. We find medium-sized PAHs (50 $\lesssim$ $N_{\rm C} \lesssim$ 90) to be normally 
   hydrogenated, while larger PAHs ($N_{\rm C} \gtrsim 90$) can be superhydrogenated, and smaller species ($N_{\rm C} 
   \lesssim$ 50) are fully dehydrogenated. In the more diffuse gas of the cavity, where the fullerene C$_{60}$ was 
   recently detected, all the studied PAHs are found to be quickly fully dehydrogenated. PAH chemical evolution exhibits 
   a complex non-linear behaviour as a function of the UV radiation field because of multiphoton events. Steady state 
   for hydrogenation is reached on timescales ranging from less than a year for small PAHs, up to 10$^4$ years for large 
   PAHs at \Av=1. Critical reactions that would need more studies are the recombination of cations with electrons, the 
   reactivity of cations with \hh and the reactivity of neutral PAHs with H.
   }
   {We developed a new model of PAH chemical evolution based on the most recent available molecular data. This model 
   allows us to rationalize the observational constraints without any fitting parameter. PAHs smaller than 50 carbon 
   atoms are not expected to survive in the NGC 7023 NW PDR. A similar conclusion is obtained for the diffuse ISM. 
   Carbon clusters turn out to be end products of PAH photodissociation, and the evolution of these clusters needs to be 
   investigated further to evaluate their impact on the chemical and physical evolution of PDRs.}


   \maketitle
%


\section{Introduction}	\label{sec:introduction}

   Polycyclic aromatic hydrocarbons (PAHs) are now widely accepted as the carriers of the ubiquitous aromatic infrared
   bands (AIBs) observed at 3.3, 6.2, 7.7, 8.6 and 11.3 $\mu$m and this has motivated a lot of studies on the properties
   of these species \citep[see][for a recent compilation of papers on this subject]{joblin_pahs_2011}. Interstellar PAHs 
   can contain up to 20\% of the cosmic carbon and are the smallest dust particles by size.
   As such, they play a major role in the physics and chemistry of photodissociation regions (PDR) through the 
   absorption of UV light \citep{joblin_contribution_1992}, the photoelectric heating \citep{bakes_photoelectric_1994} 
   and may also be involved in the formation of \hh \citep{habart_h2_2003}. 

   The energetics of PDRs is well understood in its main lines. However, it has been known for more than a decade now
   \citep[see, e.g.][]{hollenbach_photodissociation_1999} that the excitation temperatures observed for some
   molecular tracers are higher than predicted by PDR models. This issue was raised again recently by several authors 
   who compared the results of the most recent PDR modelling tools with observational data from the last generation of 
   spatial facilities ({\it Spitzer, Herschel}). These authors reported difficulties to account for the observed 
   emission in \hh and high-J CO rotational lines \citep{joblin_CO_inprep, habart_excitation_2011, goicoechea_oh_2011}. 
   \citet{habart_excitation_2011} concluded that this is due to our poor understanding of the energetics of PDRs. The 
   main heating mechanism at the surface of PDRs comes from the thermalisation of hot electrons generated by the
   photoelectric effect on PAHs and very small grains. The photoelectric efficiency depends on the grain charge and is 
   expected to decrease when PAHs are positively ionized \citep{hollenbach_photodissociation_1999}. This theoretical 
   prediction has been recently supported by the analysis of observations (\citeauthor{okada_probing_2012}, submitted).

   The question of the exact evolution of PAH properties with physical conditions has been addressed recently by
   a combination of observations and  theoretical modelling. By use of the Infrared Space Observatory (ISO)
   and the {\it Spitzer} telescope, the spatial  evolution of the mid-IR spectra was observed in several regions.
    \citet{rapacioli_spectroscopy_2005} and 
   \citet{berne_analysis_2007} used global decomposition methods to analyse the data and proposed an evolutionary 
   scenario in which a population of very small grains evaporates under the action of UV photons, releasing neutral free 
   PAHs, which are subsequently photoionized. In addition, the recent detection of fullerene C$_{60}$ in reflection
   nebulae \citep{sellgren_c60_2010} may be interpreted as a new step in this evolutionary scenario \citep{berne_formation_2012}.

   In parallel with these observational studies, several models were developed to describe the chemical evolution of PAHs. 
   \citet{bakes_photoelectric_1994} focused on the charge evolution in order to compute the contribution of PAHs to the
   gas heating by the photoelectric effect. A more comprehensive model was proposed by 
   \citet{allain_photodestruction_1996-1}, which includes the photodestruction of PAHs exposed to
   UV photons, and provides insights into the stability of interstellar PAHs, as a function of their size. 
   \citet{le_page_hydrogenation_2001} proposed a new model dedicated to the charge and hydrogenation states of PAHs in 
   the diffuse interstellar medium (ISM). Based on these models, \citet{visser_pah_2007} built a comprehensive model of 
   PAH evolution in protoplanetary disks, which includes the charge and hydrogenation states. The authors also 
   described the photodestruction of the skeleton of PAHs including multiphoton events that are frequent in these 
   high-UV irradiation conditions.
   
   In this paper, we present a new model that is inspired from \citet{le_page_hydrogenation_2001}, 
   and includes multiphoton events for all photodissociation processes. This model takes advantage of the latest
   available molecular data from both theoretical calculations \citep{malloci_theoretical_2007,malloci_dehydrogenated_2008} 
   and experimental measurements \citep{joblin_coro-h_inprep, biennier_laboratory_2006, betts_gas-phase_2006}. We 
   apply our model to the prototypical north-west PDR (hereafter NW PDR) of NGC 7023 in an attempt to (i) identify 
   key processes in the evolution of the charge and hydrogenation states of interstellar PAHs in order to provide 
   guidelines for fundamental studies and (ii) provide a quantitative description of the evolution of the PAH population, 
   which is necessary for modelling the role of PAHs in the physics and chemistry of 
   PDRs. Furthermore, the obtained results could guide the spectroscopic identification of individual PAHs for 
   instance by matching some of the diffuse interstellar bands \citep[see, for example,][]{salama_polycyclic_2011}.
   This motivates additional calculations on the hydrogenation and charge states of PAHs using a grid of conditions that 
   include typical conditions for the diffuse ISM.

   This work is organised as follows. The studied PDR and its physical conditions are presented in
   Sect.~\ref{sec:APenvironment}. In Sect.~\ref{sec:PCprocess}, molecular processes relevant to this study are reviewed 
   and discussed. Our numerical model is presented in Sect.~\ref{sec:NumericalFramework} and its results in 
   Sect.~\ref{sec:results}. The sensitivity of the results, both to molecular data and astrophysical conditions, are 
   investigated in Sect.~\ref{sec:sensitivity}. Consequences of our results on the direct detection of individual PAH 
   species and on the fate of carbon clusters are discussed in Sect.~\ref{sec:discussion}.


\section{The astrophysical environment}\label{sec:APenvironment}


   \begin{table}
      \begin{center}
         \caption{Input parameters of the Meudon PDR code \citep{le_bourlot_surface_2012} used to model the physical 
         and chemical structure of NGC 7023 NW PDR.}
         \label{tab:inputPDR}
         \begin{tabular}{l l c c }
            \toprule
            \multicolumn{2}{l}{Parameters}                        & Values            & Units      \\
            \midrule
            Star surface temp.                  & $T_{\rm eff}$   & 15000             & K          \\
            UV radiation field                  & $G_0$           & 2600              & Habing$^{\dagger}$ \\
            Fixed pressure                      & $P$             & $7\times10^6$     & K\,cm$^{-3}$ \\
            $A_{\rm V} / E_{\rm B-V}$           & $R_{\rm V}$     & 5.56              &            \\
            \midrule
                                                & $x_0$           & 4.60              & $\mu$m$^{-1}$\\
            Extinction curve with               & $\gamma$        & 1.36              & $\mu$m$^{-1}$\\ 
            Fitzpatrick \& Massa                & $c_1$           & 0.80              &            \\ 
            parameters                          & $c_2$           & 0.32              & $\mu$m\phantom{$^{-1}$}     \\ 
            for HD 200775                       & $c_3$           & 3.09              & $\mu$m$^{-2}$\\ 
                                                & $c_4$           & 0.37              &            \\
            \midrule
            CR ionization rate                  & $\zeta$         & $5\times10^{-17}$ & s$^{-1}$ \\
            Dust minimum radius                 & $a_{\rm min}$   & $3\times10^{-7}$  & cm       \\
            Dust maximum radius                 & $a_{\rm max}$   & $3\times10^{-5}$  & cm       \\
            MRN distribution index              & $\alpha$        & 3.5               &          \\
            \bottomrule
         \end{tabular}

         \tablefoot{
          $\dagger$ The fluxes are normalized by the flux of the interstellar radiation field between 912 and 2400 $\AA$ 
          measured by \citet{habing_interstellar_1968} in the $\sim 1$ kpc neighbourhood of the Sun.
         }

      \end{center}
   \end{table}

   Our study focuses on the reflection nebula NGC 7023 that has been widely studied at many wavelengths
   \citep[e.g.][]{rogers_h_1995, fuente_filamentary_1996, gerin_co_1998, fuente_infrared_1999, berne_extended_2008,
   joblin_gas_2010}. The region is part of a small molecular cloud at~430\,pc, illuminated by the Be star HD 200775
   [RA(2000)~=~21h01m36.9s ; Dec(2000)~=~+68$^{\circ}$~09~47.8]. It has been shaped by the star formation process 
   leading to the formation of a cavity surrounded by denser filaments and clumps. Figure~1 shows the NW
   part of this region.
   
   \begin{figure}
      \begin{center}
         \includegraphics[width=0.49\textwidth, keepaspectratio]{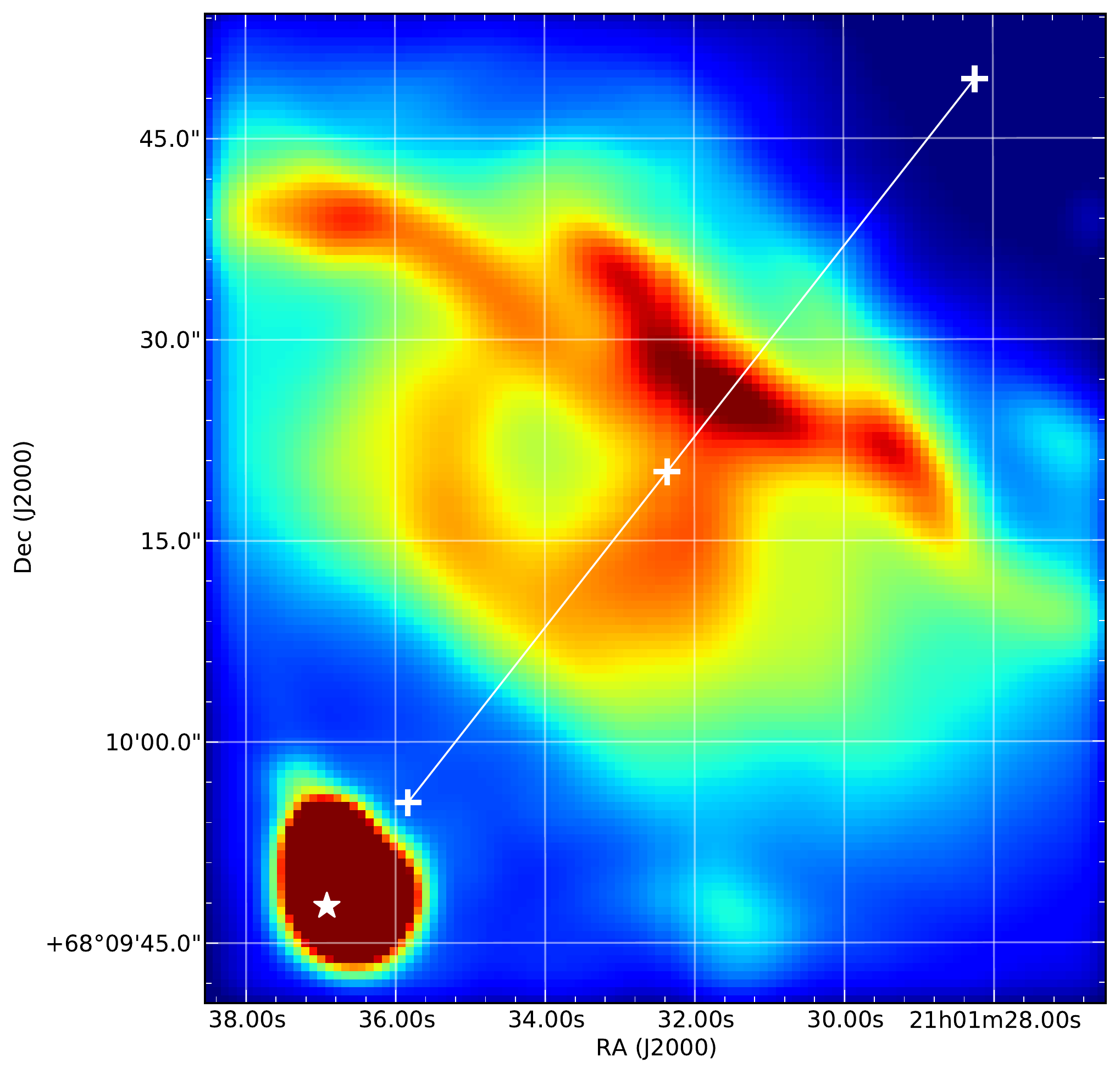}
         \caption{Map of the NW PDR of NGC 7023 observed by \textit{Spitzer}-IRAC at 8$\mic$. The white star shows the 
         location of the exciting star HD 200775. The white line shows the Star-NW cut that is studied in this paper, 
         and the white crosses indicate the positions at 10, 42 and 80\arcsec \, from the star.}
           \label{fig:map_7023}
      \end{center}
   \end{figure}

\begin{figure*}
      \begin{center}
         \includegraphics[width=0.49\textwidth, keepaspectratio]{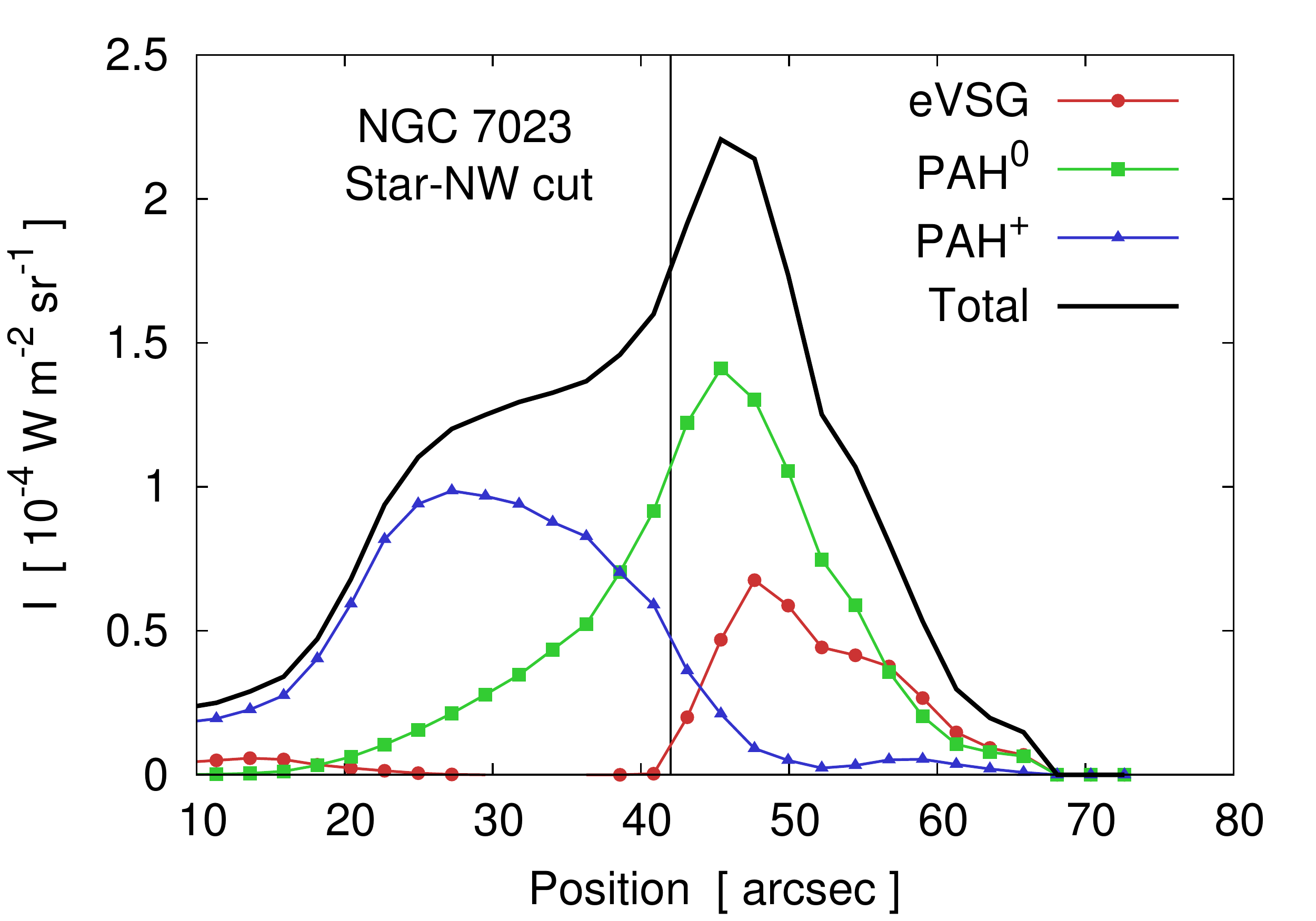}
         \includegraphics[width=0.49\textwidth, keepaspectratio]{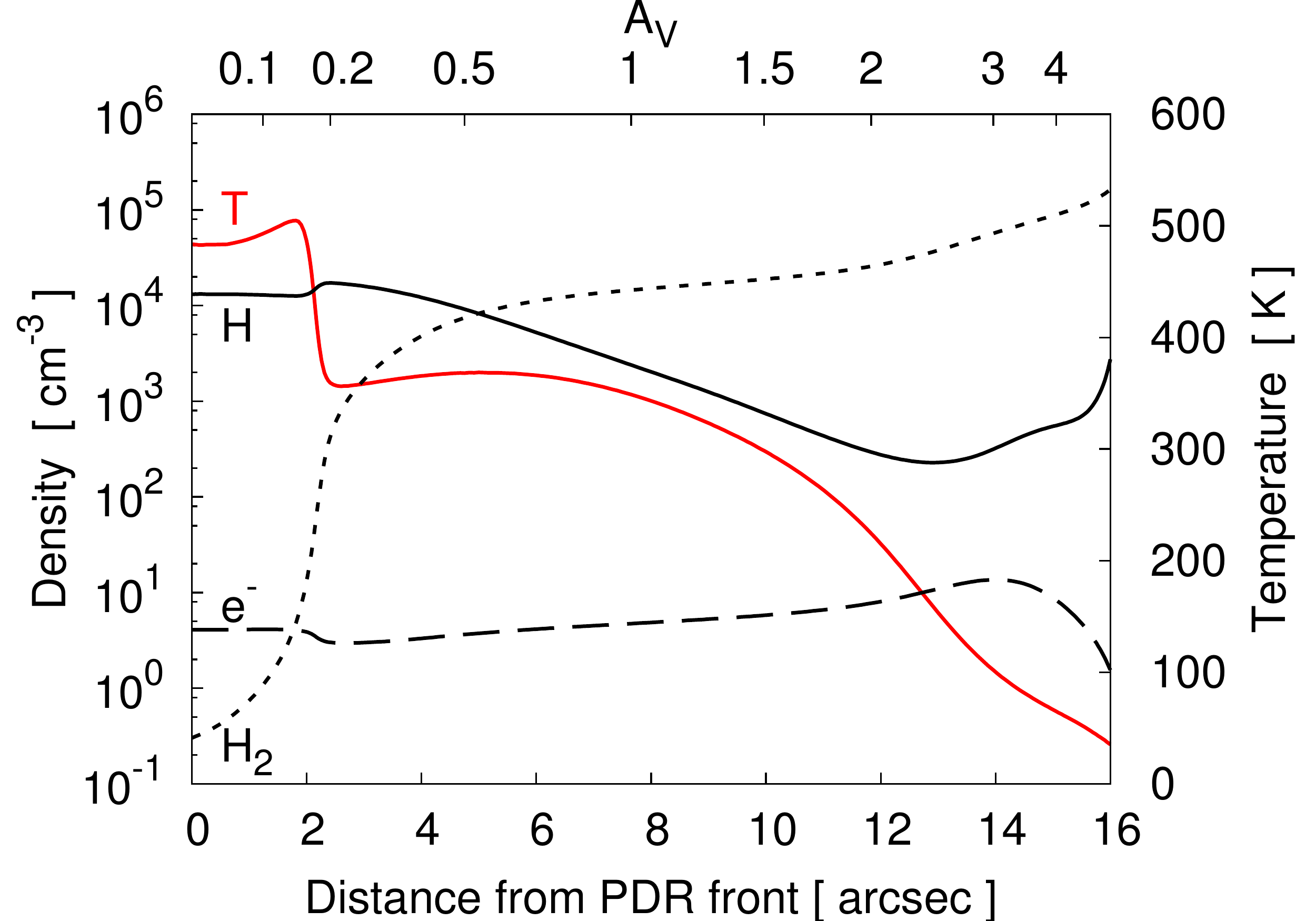}
      \end{center}
      \caption{\textit{Left:} Analysis of the AIB emission in NGC 7023 along the Star-NW cut, using the PAHTAT tool 
      \citep{pilleri_evaporating_2012}. The integrated intensity over the 5.5 - 14$\mic$ spectral  range after correction 
      from extinction (thick black), as well as the contributions from eVSGs (red), neutral PAHs (green), and PAH cations 
      (blue) are shown. The vertical line indicates the position of the PDR front, at $42\arcsec$ from the star.
      \textit{Right:} Physical and chemical structure of the NW PDR  as computed with the Meudon PDR code, assuming an 
      intensity of the UV radiation field $G_0$=2600 in Habing units at \Av=0 ($42\arcsec$ from the star), and a constant 
      pressure of $7\times10^6$ K\,cm$^{-3}$.}
      \label{fig:7023obs}
   \end{figure*}

   \citet{pilleri_evaporating_2012} used their spectral fitting tool  PAHTAT to analyse, in this region, the AIB 
   emission that was recorded by the Infrared Spectrograph (IRS) on-board {\it Spitzer}. They derived the respective 
   contributions of PAH cations (PAH$^+$), neutral PAHs (PAH$^0$) and evaporating very small grains (eVSG). 
   Figure~\ref{fig:7023obs} reports these values along the Star-NW cut shown in Fig.~\ref{fig:map_7023}. 
   In the following, we focus on two regions of the nebula along the Star-NW cut: the cavity, where the emission of 
   fullerene C$_{60}$ is detected and the AIB signal disappears \citep{sellgren_c60_2010, berne_formation_2012} and the 
   well-studied NW PDR, where the AIB emission is intense.
   
   In order to provide a realistic description of the spatial evolution of physical conditions (density and gas 
   temperature) and chemical abundances in the PDR, we used the Meudon PDR code \citep[version 1.4.3, rev. 608,][]
   {le_bourlot_surface_2012}. The input parameters are summarized in Tab.~\ref{tab:inputPDR}. The illuminating star was 
   modelled using a synthetic stellar atmosphere spectrum from \citet{castelli_new_2004} with an effective temperature 
   of 15000 K (solar elemental abundances and log(g)=4.0) and an integrated UV intensity of 2600 in units of Habing 
   \citep{habing_interstellar_1968} at $42\arcsec$ from the star \citep{joblin_gas_2010}. The UV spectrum measured with 
   IUE and the photometric U, B, V \citep{mermilliod_general_1997} and J \citep{skrutskie_two_2006} bands were combined 
   with this stellar atmosphere to determine the Fitzpatrick and Massa parameters of the extinction curve along the line
   of sight toward HD 200775 \citep{fitzpatrick_analysis_2005}. These parameters are used for radiative transfer 
   calculations in the PDR code. For the NW PDR, we computed an isobaric model with a thermal pressure of $7 \times 
   10^{6}$ K\,cm$^{-3}$, which provides a density profile in reasonable agreement with the profile derived by \citet
   {pilleri_evaporating_2012} from the {\it Spitzer} mid-IR observations, and also with the C$^+$ emission at $158\,\mic$ 
   measured with HIFI/{\it Herschel} at this location \citep{joblin_gas_2010}. For the cavity, we computed isochore 
   models for three positions at 10, 20 and $30\arcsec$ from the star. We followed \citet{berne_formation_2012} and 
   considered a density of hydrogen nuclei $\nH=150$ cm$^{-3}$ and a local UV radiation field that is derived by 
   applying a geometrical dilution factor.
   
   The resulting physical conditions and chemical abundances are shown in Fig.~\ref{fig:7023obs} as a function of the 
   position along the Star-NW cut for the NW PDR. As the extinction is negligible in the cavity, the variations of 
   the radiation field are dominated by the geometrical dilution which cannot be taken into account using the 
   plane-parallel geometry of the Meudon PDR code. Therefore, for each position in the cavity, we ran a 
   different model and checked that the extinction was negligible along distances on the order of $10\arcsec$. In the 
   following, we use the local conditions obtained at \Av=10$^{-4}$ in each cavity model, as summarized in 
   Tab.~\ref{tab:CondCavite}, and compute the evolution of PAHs in this region. In the next section, we present the 
   physical and chemical processes that drive the evolution of PAH charge and hydrogenation states.

   \begin{table}
      \begin{center}
         \caption{Local physical conditions in the cavity of NGC 7023.}
         \label{tab:CondCavite}
         \begin{tabular}{c c c c c}
            \toprule
            d & T & n(H) & n(\hhp) & n(e$^-$) \\
            ~[ $\arcsec$ ]~ & [ K ] & [ cm$^{-3}$ ] & [ cm$^{-3}$ ] & [ cm$^{-3}$ ] \\
            \midrule
            10 & 103 & 150 & $6.9 \times 10^{-8}$ & $3.7 \times 10^{-2}$ \\
            20 & 129 & 150 & $6.8 \times 10^{-7}$ & $3.9 \times 10^{-2}$ \\
            30 & 150 & 150 & $3.3 \times 10^{-6}$ & $4.0 \times 10^{-2}$ \\
            \bottomrule
         \end{tabular}
      \end{center}
   \end{table}
      
\section{Physical and chemical processes}	\label{sec:PCprocess}
   

   In the following, we focus on the evolution of the charge and hydrogenation states of a few PAHs, namely coronene 
   ($N_{\rm C}$=24), circumcoronene ($N_{\rm C}$=54), circumovalene ($N_{\rm C}$=66) and circumcircumcoronene 
   ($N_{\rm C}$=96). These molecules have compact geometries and are therefore expected to be the most stable for a 
   given size. Furthermore, coronene is the largest PAH species whose chemical properties have been studied 
   experimentally \citep{bierbaum_pahs_2011}.
   
   We review here our current knowledge on the main processes that drive the evolution of these PAHs, and justify our 
   choices when facing the lack of available molecular data. We do not consider the gain or loss of C-atoms in this 
   section as (i) reactivity with gas-phase species such as C$^+$ and O is expected to be much slower than the 
   processes related to charge and hydrogenation 
   states \citep{le_page_hydrogenation_2001}, and (ii) large and compact PAHs were shown to lose primarily all their 
   hydrogen atoms before enduring C-loss when submitted to photodissociation \citep{ekern_photodissociation_1998}. 
   
   In the following, we define the hydrogenation state as the number $N_{\rm H}$ of H-atoms bound to the carbon skeleton 
   of the molecule. We disentangle between 
   (i) normally hydrogenated PAHs, containing the maximum number $N_{\rm H}^0$ of H-atoms bound to peripheral C-atoms 
   without breaking aromaticity, 
   (ii) dehydrogenated PAHs that have lost one or several H-atoms compared to the normally hydrogenated structure 
   ($N_{\rm H} < N_{\rm H}^0$), and
   (iii) superhydrogenated PAHs with more H-atoms than in the normally hydrogenated structure ($N_{\rm H} > N_{\rm H}^0$), 
   leading to some aliphatic bonds. In practice, we will consider only species with $N_{\rm H} = N_{\rm H}^0 +1$ since 
   there is no quantitative data on more hydrogenated species.

\subsection{Absorption of UV-visible photons}\label{sigabs}

   The UV-visible absorption cross-sections of PAHs containing a few tens of C-atoms have been investigated experimentally 
   \citep{verstraete_ionization_1990, joblin_contribution_1992, ruiterkamp_spectroscopy_2002}. Larger species are 
   difficult to handle experimentally and their cross-sections are more easily obtained with theoretical methods, like 
   density functional theory \citep{malloci_electronic_2004}. One general result of these studies is that UV-visible 
   absorption cross-sections scale with the number of carbon atoms in PAHs.
   
   In this work, we use the cross-sections from the theoretical spectral database of polycyclic aromatic hydrocarbons 
   \citep{malloci_theoretical_2007}, which provides us with a consistent set of data. Since there are no calculations 
   for circumcircumcoronene (\ch{96}{24}),  we scaled,  according to the number of C-atoms, the absorption cross-section 
   of circumcoronene (\ch{54}{18}), the largest  studied PAH with similar symmetry.

\subsection{Photoionization}\label{photoion}
   
   The photoionization cross-sections of pyrene (\ch{16}{10}) and coronene (\ch{24}{12}) were measured by 
   \citet{verstraete_ionization_1990}, who found that they are well parametrised by the number of C-atoms and the 
   ionization potential (IP). We make use of the formula proposed by \citet{le_page_hydrogenation_2001}:
   \begin{equation}
      \label{eq:IonYield}
      \sigma_{\rm ion} = \sigma_{\rm UV} \times Y_{\rm ion} = \sigma_{\rm UV} \times a \times \exp\{-b [ c (E_{\rm exc} - d) ]^4\}
   \end{equation}
   where the UV-visible absorption cross-section $\sigma_{\rm UV}$ is the one discussed in the previous section and 
   accounts for the evolution with the number of C-atoms. The ionization yield $Y_{\rm ion}$ accounts for the effect of 
   the IP. The constants are taken from  \citet{le_page_hydrogenation_2001}: $a=0.8$, $b=0.00128$, $c=(14.89-$IP$_{\rm coronene})
   /(14.89-$IP$_{\rm PAH})$ where the ionization potential of coronene (IP$_{\rm coronene}$) and of the studied PAH 
   (IP$_{\rm PAH}$) are expressed in eV, and d=14.89 eV. For coronene, circumcoronene and circumovalene, we used the IP 
   computed by \citet{malloci_theoretical_2007} (respectively 7.02, 6.14 and 5.71 eV). For circumcircumcoronene, we 
   assumed an IP of 5.68 eV, consistent with the parametrisation proposed by \citet{ruiterkamp_pah_2005} in their Tab.~B.1 
   for molecules with a D$_{6h}$ geometry.

\subsection{Photodissociation}\label{photodiss}

   As discussed by \citet{allain_photodestruction_1996}, after the absorption of a UV-visible photon, a PAH can undergo 
   several transformations involving a wealth of processes: it can lose an electron or a fragment (H, \hhp, \ch{2}{2}, ...), 
   or relax energy by radiating IR photons or by luminescence. \citet{allain_photodestruction_1996} showed that the 
   branching ratios between the different pathways vary with the internal energy and the size of PAHs. For compact PAHs 
   as studied here, the main processes apart from photoionization, are H-loss and IR photon emission. In the following, 
   we therefore describe the photophysics of PAHs with a few simplified steps: (i) absorption of a UV-visible photon of 
   energy $h\nu$ either leads to ionization with the branching ratio $Y_{\rm ion}(h\nu)$, or increases, by the energy 
   $h\nu$, the internal energy of the PAH, which is then found in a vibrationally excited state of the electronic ground 
   state; (ii) emission of  IR photons at the rate $\kIR(E)$, with $E$ the internal energy of the PAH; (iii) 
   fragmentation by H-atom lost at the rate $\kdiss(E)$. In this section, we present how we determined these two rates.

   \subsubsection{IR relaxation}\label{kIR}

   \begin{figure*}
      \begin{center}
         \includegraphics[width=0.49\textwidth, keepaspectratio]{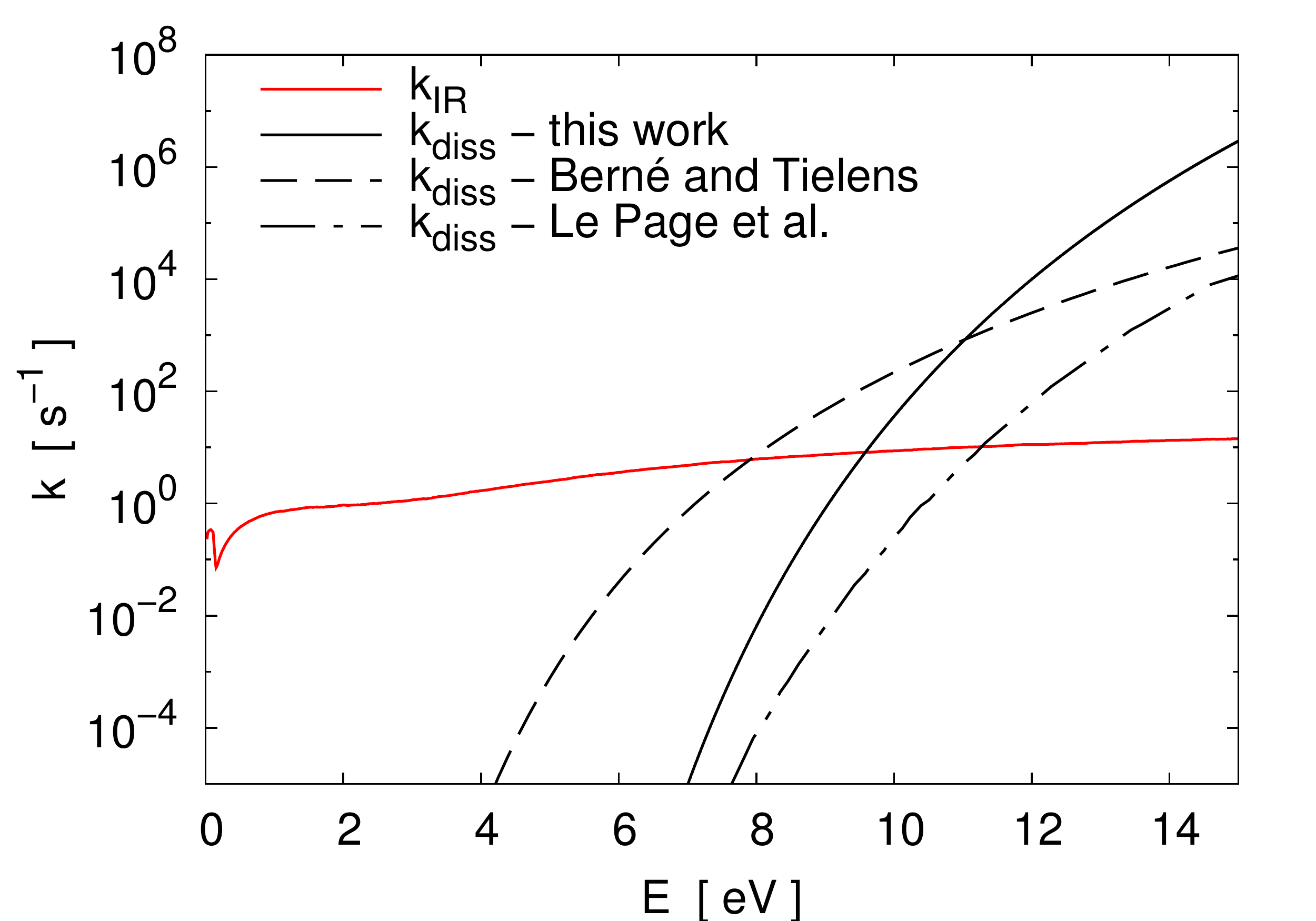} \hfill
         \includegraphics[width=0.49\textwidth, keepaspectratio]{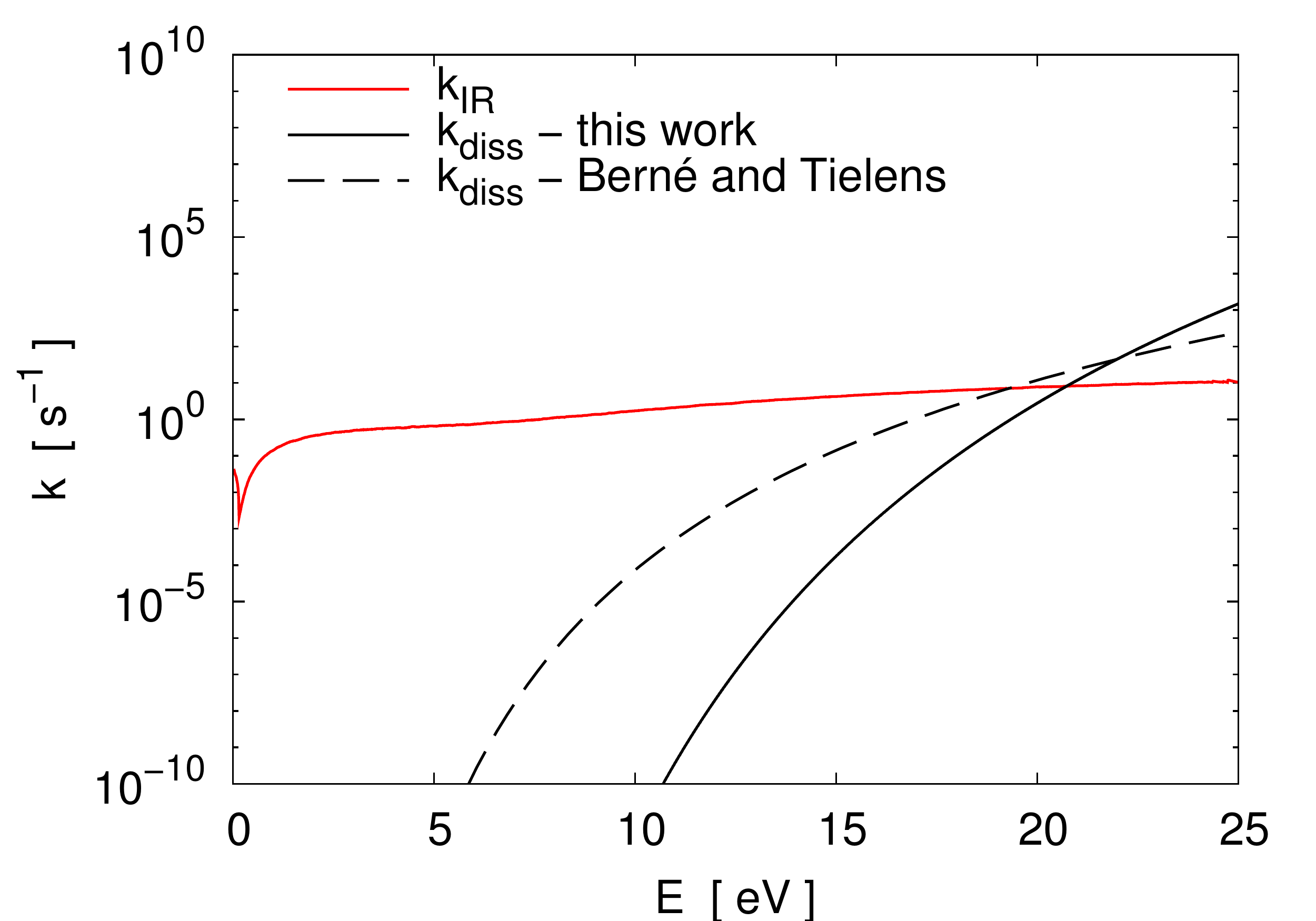}
      \end{center}
      \caption{Rates for IR emission and dissociation as a function of the internal energy for coronene \ch{24}{12} 
      (left) and circum\-ovalene \ch{66}{20} (right). For the typical energies of UV photons in PDRs, our results 
      predict dissociation rates that are much lower  than in \citet{berne_formation_2012}, 
      and much higher than in \citet{le_page_hydrogenation_2001}.}
      \label{fig:kdiss}
   \end{figure*}

   Several approaches have been proposed to compute the IR photon emission rate of PAHs, from the simple assumption that
   $\kIR \approx 10^2$ s$^{-1}$ \citep{herbst_how_1990}, up to a detailed Monte-Carlo kinetics model using a microcanonical
   formalism by \citet{joblin_calculations_2002}. The latter approach requires the full knowledge of the vibrational mode
   frequencies and associated Einstein coefficients. The values of $\kIR(E)$ are obtained by averaging over typically
   10000 trajectories. For coronene, circumcoronene and circumovalene, we used this method and the vibrational properties
   from the theoretical spectral database of PAHs \citep{malloci_theoretical_2007}. As can be seen on Fig.~\ref{fig:kdiss},
   the dependence of $\kIR$ on PAH size is not strong. For circumcircumcoronene, we used the same values of $\kIR$ as for
   circumcoronene.
   
   In addition to the determination of $\kIR(E)$, we extended the method of \citet{joblin_calculations_2002} to derive 
   the mean energy $\eIR(E)$ of the IR photons emitted as a function of the internal energy of PAHs, so that one can 
   compute the IR cooling rate as a function of the internal energy of the molecule as:
   \begin{equation}
      \label{eq:IRcooling}
      \frac{dE}{dt}(E) = \kIR(E) \times \eIR(E)
   \end{equation}

   Since the values and shape of $\kIR(E)$ were shown to very weakly affect the results of PAH evolution modelling 
   \citep{le_page_hydrogenation_2001}, both values $\kIR(E)$ and $\eIR(E)$ were computed only for cations in normal 
   hydrogenation state, and used for all charge and hydrogenation states.

   \subsubsection{Dissociation}\label{kdiss}

   Photodissociation of PAHs has been experimentally and theoretically studied by several teams. \citet{ekern_photodissociation_1998}
   qualitatively studied PAH photostability and showed that larger and more compact species are more resistant to 
   UV-visible photons, in agreement with previous results \citep{jochims_size_1994}. The same team determined the 
   branching ratios of the dissociation paths of fluorene cation (\chp{13}{10}{+}) showing that the successive loss of 
   H-atoms widely dominates the other paths like H$_2$-loss or \ch{2}{2}-loss \citep{dibben_photodissociation_2001,
   szczepanski_theoretical_2001}. The works of \citet{jochims_size_1994}, \citet{ekern_photodissociation_1998} and 
   \citet{banisaukas_photodissociation_2004} strongly suggest that this behaviour is even more pronounced for larger 
   PAHs. The photodissociation rates of naphthalene (\chp{10}{8}{+}), anthracene (\chp{14}{10}{+}), phenanthrene 
   (\chp{14}{10}{+}) and pyrene (\chp{16}{10}{+}) have been quantitatively determined by Lifshitz and co-workers 
   \citep{ho_c-h_1995,ling_time-dependent_1998,ling_time-dependent_1995} who combined experimental results with the
   Rice-Ramsperger-Kassel-Marcus (RRKM) statistical theory. They showed, in addition, that once a first H-atom is lost, 
   the loss of a second H-atom is faster \citep{ling_time-dependent_1998}. Based on these results, \citet
   {le_page_hydrogenation_2001} used a simplified version of RRKM in order to compute the dissociation rates of large 
   PAHs, bearing up to 200 C-atoms.
   
   The coronene cation (\chp{24}{12}{+}) has been more specifically studied by \citet{joblin_coro-h_inprep} with the 
   Fourier transform-ion cyclotron resonance trap (FT-ICR) experiment PIRENEA. They confirmed that full dehydrogenation 
   occurs before any carbon loss. In addition, the partially dehydrogenated cations bearing an odd number of H-atoms are 
   found to photodissociate much faster than those bearing an even number of H-atoms, consistently with the results of 
   \citet{ling_time-dependent_1998} on smaller species.

   The results of \citet{jochims_size_1994} and \citet{ling_time-dependent_1998} are well explained by statistical 
   theories. This justifies the use of statistical approaches in previous PAH models \citep{leger_photo-thermo-dissociation._1989,
   allain_photodestruction_1996,le_page_hydrogenation_2001,visser_pah_2007}. We chose to follow the approach described 
   in \citet{leger_photo-thermo-dissociation._1989} and \citet{boissel_fragmentation_1997}, based on the Laplace 
   transform of the Arrhenius law:
   \begin{equation}
      \label{eq:kdiss}
      \kdiss(E) = A_{\rm diss} \frac{\rho(E-E_0)}{\rho(E)}
   \end{equation}
   where $E$ is the internal energy of the PAH, $A_{\rm diss}$ has the dimension of a frequency, $E_0$ is the 
   dissociation energy and $\rho(E)$ is the vibrational density of states of the PAH. This very simple approach
   was shown to provide satisfying results when compared to more sophisticated methods, like RRKM theory
   \citep{barker_monte_1983}.

   The two parameters $A_{\rm diss}$ and $E_0$ were discussed for the different hydrogenation states of coronene by
   \citet{joblin_coro-h_inprep}. The authors show that assuming a dissociation energy $E_0$=4.8 eV for the C-H bond of 
   all PAHs bearing an even number of H-atoms, they manage to account for the experimental results for all these species 
   by using a single value of $A_{\rm diss}=6.8\times10^{17}$ s$^{-1}$ that was derived from the photoion appearance
   potential measured by \citet{jochims_size_1994} on \ch{24}{12}. In addition, the dissociation of PAHs bearing an odd 
   number of H-atoms is well described with the same constant $A_{\rm diss}$ by assuming $E_0=3.2$ eV. Since there 
   is no data available for the other species, we use, in the following, the same values of $A_{\rm diss}$ and 
   $E_0$ for the four PAHs considered in this work, in their cationic and neutral states. Therefore, the evolution of 
   $\kdiss$ with molecular size is driven by the dependence of the vibrational densities of states on size, which 
   basically reflects the evolution of the number of degrees of freedom. We computed the exact quantum harmonic 
   vibrational densities of states with the method of Beyer \& Swinehart \citep{stein_accurate_1973}.
   
   Data concerning the dissociation of superhydrogenated PAHs are scarce. They are thought to be more easily dissociated
   than molecules in a lower hydrogenation state \citep[][and references therein]{le_page_hydrogenation_2001}. Therefore, 
   in the following, we use the dissociation rate of the corresponding PAH missing a single H-atom as a lower limit for 
   their dissociation rate (e.g. $\kdiss$(\ch{24}{13}) = $\kdiss$(\ch{24}{11}) ), leading to an upper limit on the 
   abundances of superhydrogenated species in our results.
   
   Compared to other studies, our work provides significantly different values. The new experimental constraints on 
   coronene cations lead to much more fragile species than predicted by \citet{le_page_hydrogenation_2001} (see 
   Fig.~\ref{fig:kdiss}). On the opposite, the more empirical method that is presented in \citet{tielens_book_2005} and 
   used in \citet{berne_formation_2012}, leads to very different values and shapes of $\kdiss(E)$, and a lower stability 
   compared to our values. We expect our dissociation rates to provide better results because they are based on 
   experimental data on larger PAHs than considered in previous studies.

\subsection{Reactivity with hydrogen}\label{kH}

   \begin{table*}
      \begin{center}
         \caption{Summary of the measured and inferred reaction rates of atomic and molecular hydrogen with coronene 
         cations in their various hydrogenation states.}
         \label{tab:hydro-coro}
         \begin{tabular}{l c c l c}
            \hline
            Reaction                  &    $k$ [$\rm{cm}^3\,\rm{s}^{-1}$]  & Ref.  & Comments & Evolution with $N_{\rm C}$\\
            \hline 
            \chp{24}{12}{+}   + H     & $1.4 \times 10^{-10} $      &  (a)       & FA-SIFT & $y/x$ \\
            \chp{24}{12-2n}{+}   + H  & $1.4 \times 10^{-10} $      &  (b,c,d)   & extrapolation from \chp{10}{6}{+} and \chp{16}{8}{+}  & $y/x$ \\
            \chp{24}{12-2n+1}{+} + H  & $\sim5 \times 10^{-11} $    &  (b,c)     & extrapolation from \chp{10}{7}{+} & constant \\
            \chp{24}{12+n}{+} + H     & $\sim 10^{-12} $            &  (e)       & extrapolation from \chp{6}{7}{+}, \chp{10}{9}{+} and \chp{16}{11}{+} & constant \\
            \hline 
            \chp{24}{12}{+}   + \hh    & $<5 \times 10^{-13} $      &  (a)       & FA-SIFT & constant \\
            \chp{24}{12-2n}{+}   + \hh & $<5 \times 10^{-13} $      &  (b)       & extrapolation from \chp{10}{6}{+} and \chp{16}{8}{+} & constant \\
            \chp{24}{12-2n+1}{+} + \hh & -                          &  -         & - & - \\
            \chp{24}{12+n}{+} + \hh    & -                          &  -         & - & - \\
            \hline
         \end{tabular}

         \tablefoot{
         The type of experimental set-up used to perform the measurements is given in the "Comments" column. When 
         "extrapolation" is mentioned, the rate value was extrapolated from measurements on smaller species. 
         FA-SIFT: flowing afterglow-selected ion flow tube. The last column mention the kind of variation of the rates 
         with the size of PAHs that we considered in our model. When $y/x$ is mentioned, the rate varies
         proportionately to the ratio of peripheral ($y$) to total ($x$) C-atoms in the species.
         }
         \tablebib{
         (a)~\citet{betts_gas-phase_2006}; (b)~\citet{le_page_hydrogenation_2001}; (c)~\citet{le_page_gas_1999}; 
         (d)~\citet{le_page_reactions_1999}; (e)~\citet{snow_interstellar_1998}.
         }

      \end{center}
   \end{table*}

   Only few experimental studies on the reactivity of gas-phase PAHs with hydrogen have been published, and all of them 
   concern small cations bearing up to 24 C-atoms. Therefore, we evaluate first the reaction rates for coronene cations 
   in all their hydrogenation states from the available data, and then extrapolate these values for the larger PAHs 
   considered in our work. The values of reaction rates for \chp{24}{N}{+} and their origin are summarized in Tab.~\ref
   {tab:hydro-coro}.
   
   Reactions between PAHs and atomic or molecular hydrogen generally lead to the addition of one or two H-atoms to
   peripheral C-atoms of the PAH \citep[see discussion by][]{le_page_hydrogenation_2001}.
   The reactivity of PAH cations with atomic hydrogen is found to be rather fast and to decrease
   when PAH size increases: $2.3 \times 10^{-10}$ cm$^3$\,s$^{-1}$ for benzene \citep[\chp{6}{6}{+},][]{petrie_gas-phase_1992}, 
   $1.9 \times 10^{-10}$ cm$^3$\,s$^{-1}$ for naphthalene \citep[\chp{10}{8}{+},][]{snow_interstellar_1998}, $1.4 \times 
   10^{-10}$ cm$^3$\,s$^{-1}$ for pyrene \citep[\chp{16}{10}{+},][]{le_page_gas_1999}, and $1.4 \times 10^{-10}$ cm$^3$\,s$^{-1}$ 
   for coronene \citep[\chp{24}{12}{+},][]{betts_gas-phase_2006}.
   
   Following \citet{le_page_hydrogenation_2001} we use, for normally hydrogenated PAH cation \chp{x}{y}{+}, a reaction
   rate that scales with the reaction rate of coronene, according to $k_{x,y}=(y/x)\,k_{\rm coronene}$. This expression 
   is in reasonable agreement with available experimental results considering a typical experimental uncertainty of 50\% 
   \citep{betts_gas-phase_2006}. We also follow \citet{le_page_hydrogenation_2001} to estimate reaction rates for 
   dehydrogenated and superhydrogenated PAHs, assuming that (i) dehydrogenated coronene cations missing an even number 
   of H-atoms react with the same rate than the normally hydrogenated coronene cation, as inferred from the behaviour of
   naphthalene and pyrene dehydrogenated cations; (ii) dehydrogenated coronene cations missing an odd number of H-atoms
   react with a lower rate of $\sim5 \times 10^{-11}\,\rm{cm}^3\,\rm{s}^{-1}$ similarly to \chp{10}{7}{+}; (iii)
   super-hydrogenated coronene has a lower rate of ${\sim10^{-12}\,\rm{cm}^3\,\rm{s}^{-1}}$, as suggested by measurements
   on smaller PAHs \citep{snow_interstellar_1998}. The reaction rates for the other PAH species are derived by using a
   similar scaling for even-dehydrogenated PAHs as for normally hydrogenated PAHs. No evolution of reaction rates with
   size is considered for odd-dehydrogenated PAHs and superhydrogenated PAHs.

   The reaction rates of PAH cations with molecular hydrogen were determined to be below the experimental detection 
   threshold, which led to an upper limit of $\sim 5 \times 10^{-13}$ cm$^3$\,s$^{-1}$ for the reaction rate of 
   \chp{24}{12}{+} + \hh \citep{betts_gas-phase_2006}. We therefore calculate upper and lower limits to the abundances
   of PAH hydrogenation states by assuming either that PAHs do not react with \hhp, or that PAH cations react with \hh 
   with a reaction rate of $5 \times 10^{-13}$ cm$^3$\,s$^{-1}$ leading to the addition of two H-atoms to peripheral 
   sites.   

   The reactivity of neutral PAHs with atomic hydrogen is unknown. \citet{mebel_theoretical_1997} theoretically showed 
   the absence of a potential barrier for the reaction \ch{6}{6} + H, suggesting that reaction rates may be high. 
   However, \citet{le_page_hydrogenation_2001} found that the results were not affected when including a reaction rate 
   of neutral PAHs with atomic hydrogen of $10^{-10}$ cm$^3$\,s$^{-1}$. In our standard model, we neglect the reactivity 
   of neutral PAHs, and evaluate afterwards the uncertainty resulting from this choice, arbitrarily assuming reaction 
   rates of $3 \times 10^{-11}$ cm$^3$\,s$^{-1}$ at 300 K for radical neutral PAHs (odd number of H-atoms), and $3 
   \times 10^{-13}$ cm$^3$\,s$^{-1}$ at 300 K for closed-shell neutral PAHs (even number of H-atoms).

\subsection{Recombination with electrons}\label{recomb}

%
   
   Recombination of PAH cations with electrons leads to an increase of the PAH internal energy by the value of the 
   ionization potential. The statistical approach presented in Sect.~\ref{kdiss} can then be applied to determine
   whether this extra energy is relaxed radiatively or triggers molecular dissociation. According to the rates shown in 
   Fig.~\ref{fig:kdiss} in the case of coronene, IR photon emission is at least 100 times faster than dissociation for 
   an internal energy of 7 eV, and the recombination of coronene cations with electrons is therefore expected to be 
   non-dissociative. Larger species have smaller values of their ionization potentials and are more stable than coronene.
   We can therefore neglect the dissociative channel for all the species studied here. 

   The recombination rates of PAH cations with electrons were measured for a number of small PAHs \citep
   {abouelaziz_measurements_1993, hassouna_reactions_2003, rebrion-rowe_experimental_2003, novotny_recombination_2005, 
   biennier_laboratory_2006}. All the experimental values are about one order of magnitude lower than the Spitzer law,
   which is based on the classical model of a thin conducting disk \citep{spitzer_physical_1978, verstraete_ionization_1990}. 
   However this difference decreases when the size of PAHs increases. To our knowledge, there is no published model 
   available that accurately accounts for this trend. Therefore, for the four PAHs of this study, we assume that the 
   recombination rates are comprised between the experimental rate of pyrene cations and the Spitzer law, and use the 
   average of these two values in our model. This leads to rates roughly one order of magnitude higher than those 
   proposed by \citet{le_page_hydrogenation_2001}.

\subsection{Charge exchange}\label{sec:Qexc}

   Charge exchange reactions between neutral PAHs and ions such as C$^+$, Fe$^+$ or Si$^+$ may impact the ionization
   balance of PAHs. Reactions with C$^+$ proceed at the Langevin rate \citep{canosa_reaction_1995}, and can marginally 
   affect the PAH charge state in low UV irradiation environments like the diffuse ISM \citep{wolfire_chemical_2008}. 
   Photoionization rapidly dominates the ionization balance of PAHs when the UV radiation field intensity increases. 
   Other possible partners for charge exchange are much less abundant than C$^+$ and are therefore completely negligible. 
   In the following, we do not consider charge exchange reactions. We will check the validity of this assumption in
   Sect.~\ref{sec:timescales}.

\subsection{Sum-up of our standard model}\label{sec:standard}

   \begin{table*}[t]
      \begin{center}
         \caption{Species and main parameters of the processes considered in our standard model.}
         \label{tab:model}
         \begin{tabular}{c c c c c || c c || c c | c c }
               \toprule
               & $N_{\rm{C}}$ & $N_{\rm{H}}^0$ & $q$& $N_{\rm{H}}$    & IP     & $\alpha$ (300 K)     & $E_0$  & $A_{\rm{diss}}$                       & $k_{+\rm{H}}$ (300 K)     & $k_{+\rm{H_2}}$         \\
               &              &                &    &                 & [eV]   & [cm$^{3}$\,s$^{-1}$] & [eV]   & [s$^{-1}$]                            & [cm$^{3}$\,s$^{-1}$]      & [cm$^{3}$\,s$^{-1}$]  \\
               \midrule                                                                                                                                                                                                                                  
   \multirow{6}{*}{\includegraphics[width=3\benz, keepaspectratio]{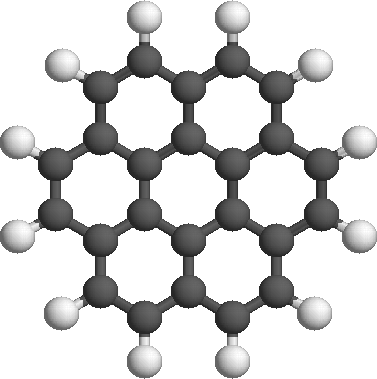}} &
   \multirow{6}{*}{24}   &\multirow{6}{*}{12}   &        & $2n$            & 7.02   & -                   & 4.8    & \multirow{6}{*}{$6.8 \times 10^{17}$}   & 0 / $3 \times 10^{-13}$  & 0                     \\
                                             &  &  &  0  & $2n+1$          & 7.02   & -                   & 3.2    &                                         & 0 / $3 \times 10^{-11}$  & 0                     \\
                                             &  &  &     & 13              & 7.02   & -                   & 3.2    &                                         & 0 / $3 \times 10^{-13}$  & 0                     \\
                                             &  &  &     & $2n$            & -      & $1.0\times10^{-5}$  & 4.8    &                                         & $1.4 \times 10^{-10}$    & 0 / $5   \times 10^{-13}$ \\
                                             &  &  &  +  & $2n+1$          & -      & $1.0\times10^{-5}$  & 3.2    &                                         & $5   \times 10^{-11}$    & 0 / $5   \times 10^{-13}$ \\
                                             &  &  &     & 13              & -      & $1.0\times10^{-5}$  & 3.2    &                                         & -                        & -                     \\
               \midrule                                                                                                                                                                                                                                  
   \multirow{6}{*}{\includegraphics[width=5\benz, keepaspectratio]{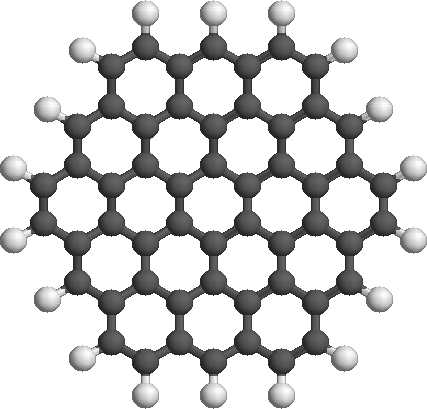}} &
   \multirow{6}{*}{54}   &\multirow{6}{*}{18}   &        & $2n$            & 6.14   & -                   & 4.8    &  \multirow{6}{*}{$6.8 \times 10^{17}$}  & 0 / $3 \times 10^{-13}$  & 0                     \\
                                             &  &  &  0  & $2n+1$          & 6.14   & -                   & 3.2    &                                         & 0 / $3 \times 10^{-11}$  & 0                     \\
                                             &  &  &     & 19              & 6.14   & -                   & 3.2    &                                         & 0 / $3 \times 10^{-13}$  & 0                     \\
                                             &  &  &     & $2n$            & -      & $1.4\times10^{-5}$  & 4.8    &                                         & $9.3 \times 10^{-11}$    & 0 / $5   \times 10^{-13}$ \\
                                             &  &  &  +  & $2n+1$          & -      & $1.4\times10^{-5}$  & 3.2    &                                         & $5   \times 10^{-11}$    & 0 / $5   \times 10^{-13}$ \\
                                             &  &  &     & 19              & -      & $1.4\times10^{-5}$  & 3.2    &                                         & -                        & -                     \\
               \midrule                                                                                                                                                                                                                                  
   \multirow{6}{*}{\includegraphics[width=6\benz, keepaspectratio]{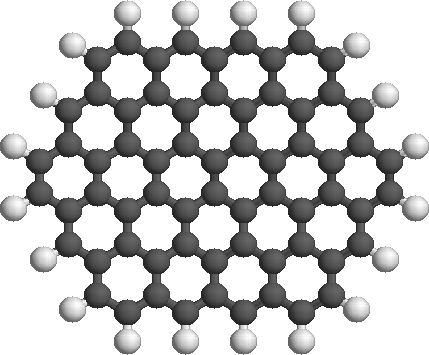}} &
   \multirow{6}{*}{66}   &\multirow{6}{*}{20}   &        & $2n$            & 5.71   & -                   & 4.8    &  \multirow{6}{*}{$6.8 \times 10^{17}$}  & 0 / $3 \times 10^{-13}$  & 0                     \\
                                             &  &  &  0  & $2n+1$          & 5.71   & -                   & 3.2    &                                         & 0 / $3 \times 10^{-11}$  & 0                     \\
                                             &  &  &     & 21              & 5.71   & -                   & 3.2    &                                         & 0 / $3 \times 10^{-13}$  & 0                     \\
                                             &  &  &     & $2n$            & -      & $1.56\times10^{-5}$ & 4.8    &                                         & $8.5 \times 10^{-11}$    & 0 / $5   \times 10^{-13}$ \\
                                             &  &  &  +  & $2n+1$          & -      & $1.56\times10^{-5}$ & 3.2    &                                         & $5   \times 10^{-11}$    & 0 / $5   \times 10^{-13}$ \\
                                             &  &  &     & 21              & -      & $1.56\times10^{-5}$ & 3.2    &                                         & -                        & -                     \\
               \midrule                                                                                                                                                                                                                                     
   \multirow{6}{*}{\includegraphics[width=7\benz, keepaspectratio]{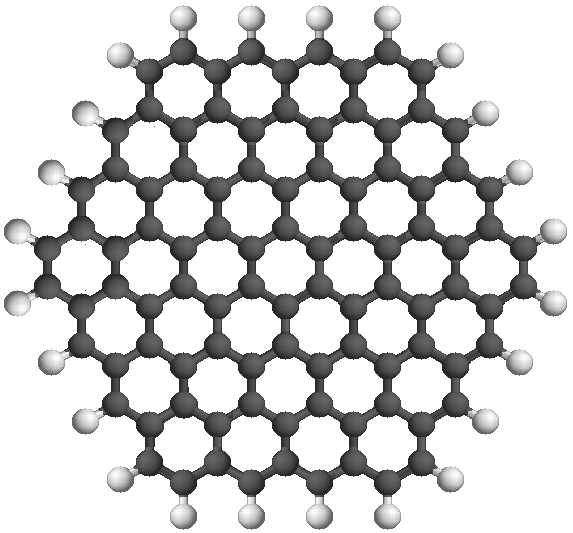}} &
   \multirow{6}{*}{96}   &\multirow{6}{*}{24}   &        & $2n$            & 5.68   & -                   & 4.8    &  \multirow{6}{*}{$6.8 \times 10^{17}$}  & 0 / $3 \times 10^{-13}$  & 0                     \\
                                             &  &  &  0  & $2n+1$          & 5.68   & -                   & 3.2    &                                         & 0 / $3 \times 10^{-11}$  & 0                     \\
                                             &  &  &     & 25              & 5.68   & -                   & 3.2    &                                         & 0 / $3 \times 10^{-13}$  & 0                     \\
                                             &  &  &     & $2n$            & -      & $1.85\times10^{-5}$ & 4.8    &                                         & $7.0 \times 10^{-11}$    & 0 / $5   \times 10^{-13}$ \\
                                             &  &  &  +  & $2n+1$          & -      & $1.85\times10^{-5}$ & 3.2    &                                         & $5   \times 10^{-11}$    & 0 / $5   \times 10^{-13}$ \\
                                             &  &  &     & 25              & -      & $1.85\times10^{-5}$ & 3.2    &                                         & -                        & -                     \\
               \bottomrule
         \end{tabular}
         
         \tablefoot{ The columns give: 
         (1) the structure of the species studied in this paper in their neutral normally hydrogenated state; 
         (2) the number $N_{\rm C}$ of C-atoms in the species; 
         (3) the number $N_{\rm H}^0$ of H-atoms in the normally hydrogenated state; 
         (4) the charge $q$ of the species; 
         (5) the number $N_{\rm H}$ of H-atoms in the species; $2n$ is for normally hydrogenated PAHs and PAHs missing 
         an even number of H-atoms, $2n+1$ is for PAHs missing an even number of H-atoms; other values are 
         superhydrogenated states;
         (6) the ionization potential IP of the species; 
         (7) the rate of recombination $\alpha$ with electrons at 300 K; 
         (8) the binding energy $E_0$ and 
         (9) the pre-exponential factor $A_{\rm{diss}}$ used in Eq.~\ref{eq:kdiss}; 
         (10) the reaction rate for H-atom addition at 300 K; 
         (11) the rate for reaction with \hhp.
         For (10) and (11), when two values are given, they correspond to lower/upper limits considered in our model.
         }
      \end{center}
   \end{table*}

   The species and processes included in our model are summarized in Tab.~\ref{tab:model}. In short, we consider here 
   four sizes of PAHs in various hydrogenation states and two charge states, namely coronene (\chp{24}{N}{0/+}), 
   circum\-coronene (\chp{54}{N}{0/+}), circum\-ovalene (\chp{66}{N}{0/+}) and circum\-circum\-coronene (\chp{96}{N}{0/+}).
   We limit the possible charge states to neutrals and monocations, excluding dications and anions for several converging 
   reasons, including (i) the lack of data for the recombination of dications with electrons and for the association of 
   neutral PAHs with electrons, and (ii) the expected weak abundances of these other charge states in the environments 
   considered here. The latter point can be checked afterwards. We consider all the hydrogenation states from fully 
   dehydrogenated species, i.e. pure carbon clusters, to superhydrogenated PAHs bearing one extra H-atom. This latter 
   choice is motivated by the lack of quantitative data for both the dissociation rates of superhydrogenated species and 
   the rates of reaction of these species with hydrogen. The dissociation rate that is used for superhydrogenated PAHs 
   should be considered as a lower limit.
   
   The charge evolution is driven by the balance between photoionization and recombination with electrons, both
   depending exclusively on the size of PAHs, in our model. Similarly, the hydrogenation state is driven by the balance
   between photodissociation and reactivity with atomic and molecular hydrogen. Photodissociation strongly depends on
   the size of the species through their vibrational density of states, and also strongly on their hydrogenation state 
   through the binding energy $E_0$, PAHs with an odd number of H being less stable than PAHs with an even number of H. 
   Dependence on the charge state is not taken into account in our model for this process. On the contrary, reactivity 
   with atomic and molecular hydrogen mainly depends on the charge and, to a lower extent, on the hydrogenation state of 
   PAHs, and more marginally on their size. Considering the lack of data for the reaction rates of PAH cations with \hh 
   and of neutral PAHs with H, we consider lower and upper values for these processes.
   
   Describing PAH evolution requires to model the processes discussed above, as well as how they vary with 
   astrophysical conditions. The UV-visible radiation field spectrum drives both photoionization and photodissociation. 
   The abundances of H, \hh and free electrons are involved through their reactions with PAHs, in which the gas 
   kinetic temperature $T$ plays a role. Considering only the influence of temperature on the collision rates, the rates 
   of recombination with electrons vary like $T^{-1/2}$ \citep[see, e.g.,][]{bakes_photoelectric_1994}, the rates of 
   reaction between two neutral species like $T^{1/2}$, and no variation with $T$ is considered for reactions between 
   charged and neutral species \citep[][and references therein]{le_page_hydrogenation_2001, herbst_chemistry_2001}.


\section{Numerical framework}  \label{sec:NumericalFramework}

\subsection{Description of the species}\label{sec:DescrSpecies}
   
   The species considered in this work are presented in Sect.~\ref{sec:standard} and Tab.~\ref{tab:model}. The evolution 
   of these species involves some processes that critically depend on the internal energy of the reactants, like 
   photodissociation processes. An explicit description of this internal energy was proposed by \citet{visser_pah_2007} 
   in the context of protoplanetary disks, where the strong UV-visible irradiation by the close illuminating star is 
   expected to favour multiphoton events, i.e. absorption of another photon before complete cooling of the PAHs.
   Previous studies on the evolution of PAHs in regions of low or moderate excitation do not describe explicitly the 
   evolution of the internal energy of PAHs \citep{allain_photodestruction_1996-1, le_page_hydrogenation_2001, 
   berne_formation_2012}. Nevertheless, a quick examination of the rates for IR photon emission ($\kIR$) and 
   dissociation ($\kdiss$, see Fig.~\ref{fig:kdiss}) reveals that for large enough species, the threshold energy $E_{\rm 
   th}$ for which $\kIR(E_{\rm th}) = \kdiss(E_{\rm th})$ falls well above the Lyman cut-off at 13.6 eV. As a consequence, 
   even in low radiation fields, dissociation is driven by multiphoton events, since this is the only way to raise the 
   internal energy of large PAHs close to or above $E_{\rm th}$. 
   
   In order to describe the evolution of internal energy, each species $X$ was divided into an ensemble of subspecies 
   $X_{i}$ (i=0, $\dots$, $N_{\rm bin}$) whose number density is the number per unit of volume of species $X_i$ bearing 
   an internal energy between $(i-1) \times \Delta E$ and $i \times \Delta E$ for (i=1, $\dots$, $N_{\rm bin}$) or no 
   internal energy for $i=0$. Thus, instead of studying the evolution of an ensemble of species, we study the evolution 
   of the histograms of their internal energy. Obviously, the accuracy of the method depends on the width of the bins, 
   $\Delta E$, that we found to be optimum for $\Delta E=0.25$ eV, which is a compromise  between accuracy and computing 
   time (see appendix~\ref{anx:binwidth}).

\subsection{Evolutionary scheme}\label{sec:EvolutionaryScheme}

   We aim at computing the time evolution of the species defined in the previous section in a specific environment. The 
   feedback of PAH evolution on their astrophysical environment is out of the scope of this work. Therefore, the densities
   of H, \hh and free electron are considered as fixed external parameters, as are the kinetic gas temperature and the 
   UV-visible flux. We compute the time evolution of each species $X$ and its internal energy histograms by considering 
   the chemical balance of each subspecies $X_{i}$:
   \begin{equation}
      \label{eq:reaction-dissociation}
      \frac{\partial [X_{i}]}{\partial t} = P_{i} - L_{i}
   \end{equation}
   where $[X_{i}]$ is the number density of the species $X_{i}$, $P_{i}$ and $L_{i}$ are the production and loss rates, 
   respectively. These two latter quantities are computed from (i) the conditions of the astrophysical environment, 
   and (ii) the rates of the processes presented in Sect.~\ref{sec:PCprocess}. Our work differs from previous studies by 
   the way we model photodissociation processes. Instead of computing an effective photodissociation rate for each 
   species $X$, we compute the production and loss rates for each subspecies $X_i$ according to the following scheme:
   \begin{eqnarray}
   X_{i} + h\nu_{_{\rm UV}} & \buildrel [1-Y_{\rm ion}(i)]k_{\rm abs}(i) \over\longrightarrow & X_{i^{\prime}} \\
   X_{i} & \buildrel k_{{\rm IR}}(i)                    \over\longrightarrow & X_{i^{\prime\prime}} + h\nu_{_{\rm IR}}\\
   X_{i} & \buildrel k_{{\rm diss}}(i)                  \over\longrightarrow & X^{-\rm H}_{0} + {\rm H}
   \end{eqnarray}
   
   \noindent where $i^{\prime} > i$, $i^{\prime\prime} < i$, and $X^{-\rm H}_{0}$ has no internal energy. The numerical 
   rates $[1-Y_{\rm ion}(i)]k_{\rm abs}(i)$, $k_{{\rm IR}}(i)$ and $k_{{\rm diss}}(i)$ are computed from the physical 
   quantities $Y_{\rm ion}(E)$, $\sigma_{\rm abs}(E)$, the UV-visible flux, $k_{{\rm IR}}(E)$ 
   and $k_{{\rm diss}}(E)$, as detailed in appendix~\ref{anx:Eint}. For the other processes, we do not take into account
   the effects of the internal energy of PAHs, which makes the numerical implementation straightforward.


\section{Results}\label{sec:results}

   \begin{figure*}
      \begin{center}
         \includegraphics[width=0.49\textwidth, keepaspectratio]{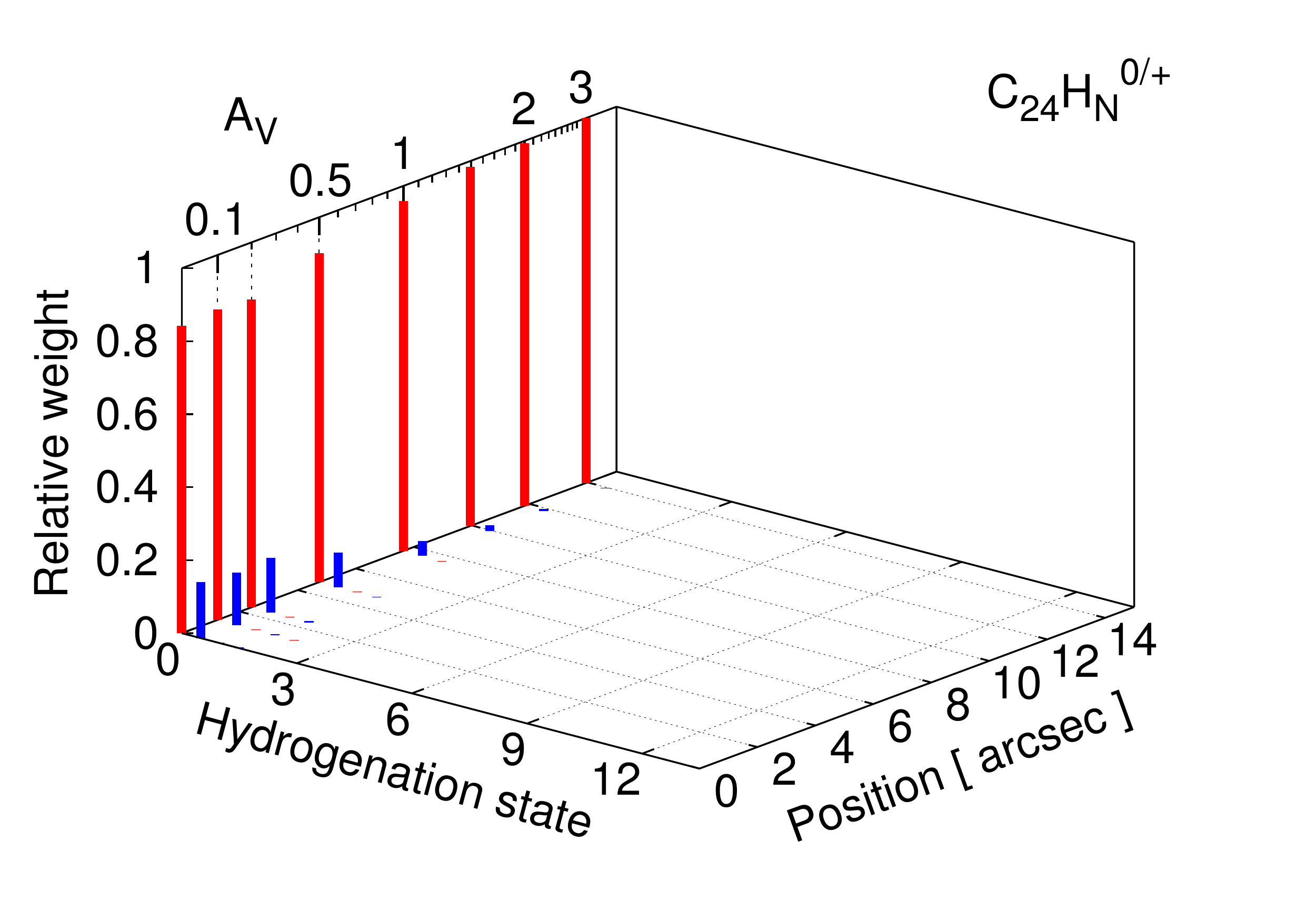} \hfill
         \includegraphics[width=0.49\textwidth, keepaspectratio]{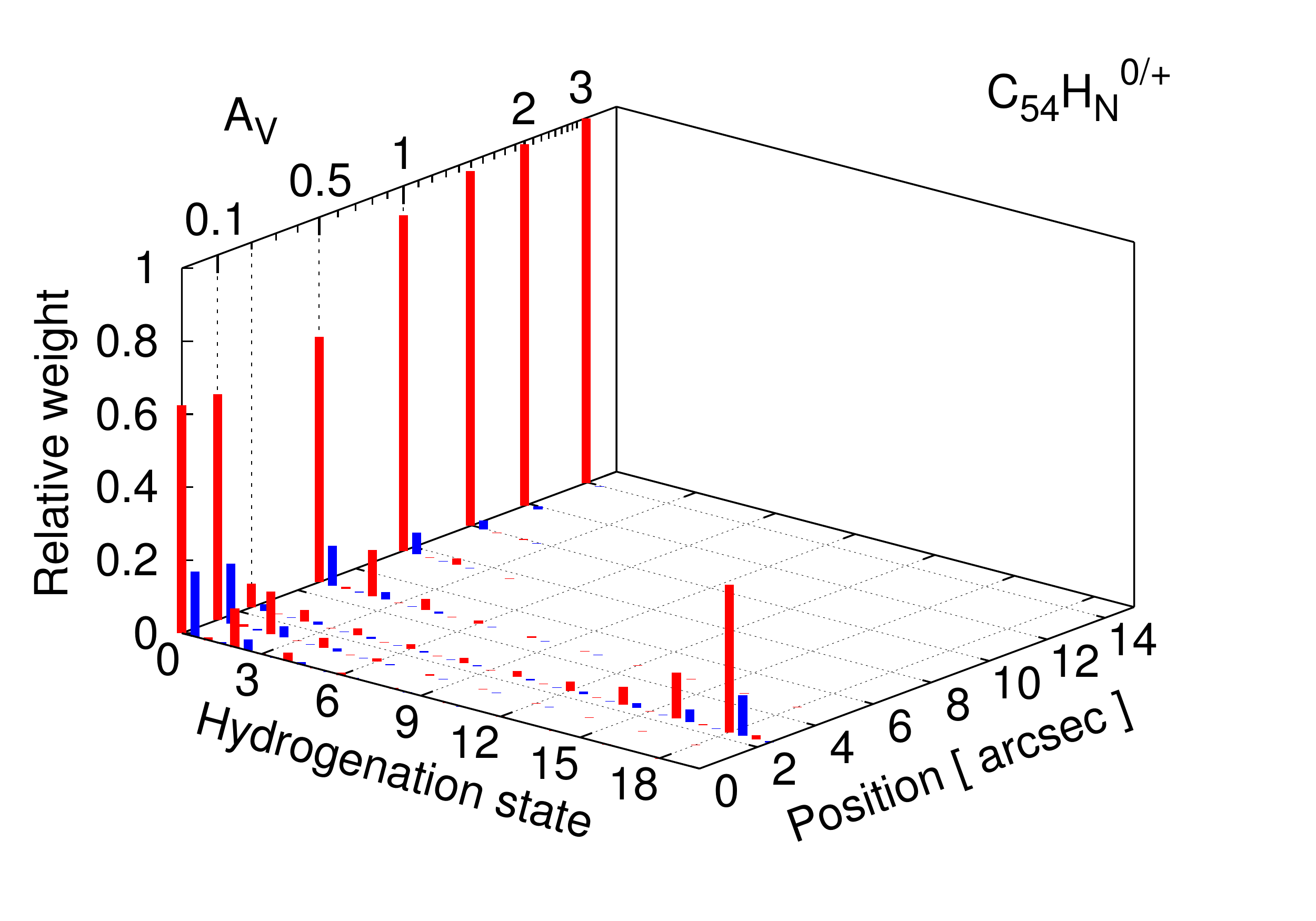}\\
         \includegraphics[width=0.49\textwidth, keepaspectratio]{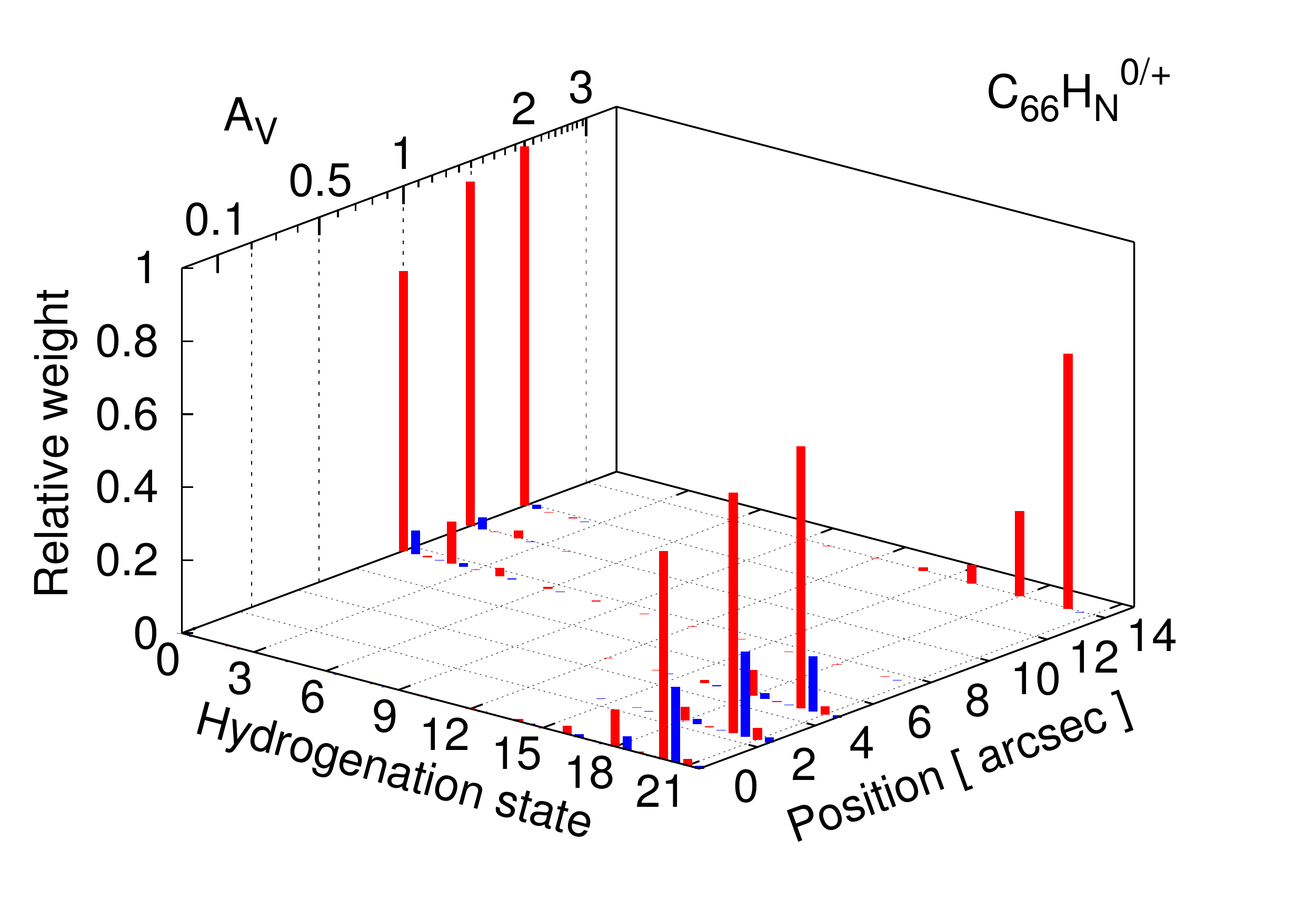} \hfill
         \includegraphics[width=0.49\textwidth, keepaspectratio]{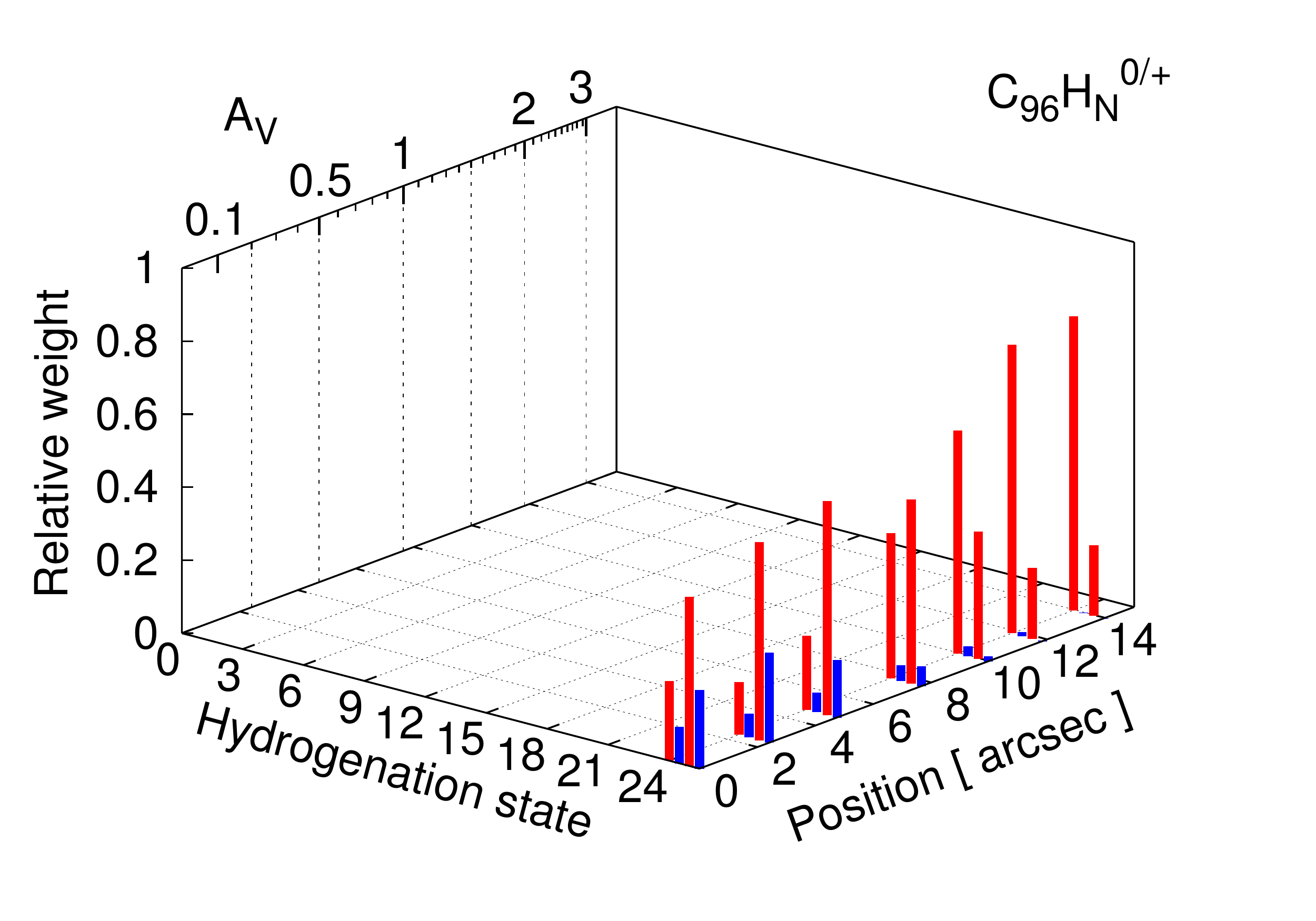}
      \end{center}
      \caption{Charge and hydrogenation states (number of H-atoms in the PAH) of coronene (\ch{24}{12}), circumcoronene 
      (\ch{54}{18}), circumovalene (\ch{66}{20}) and circum\-circum\-coronene (\ch{96}{24}) as a function of the 
      position in the NW PDR of NGC 7023, with the astrophysical conditions presented in Fig.~\ref{fig:7023obs} and with 
      $k_{\rm +H_2}=0$ and $k_{\rm +H}({\rm neutral})=0$. Neutral species are in red, cations in blue.}
      \label{fig:Results_standard}
   \end{figure*}

   \begin{figure*}
      \begin{center}
         \includegraphics[width=0.49\textwidth, keepaspectratio]{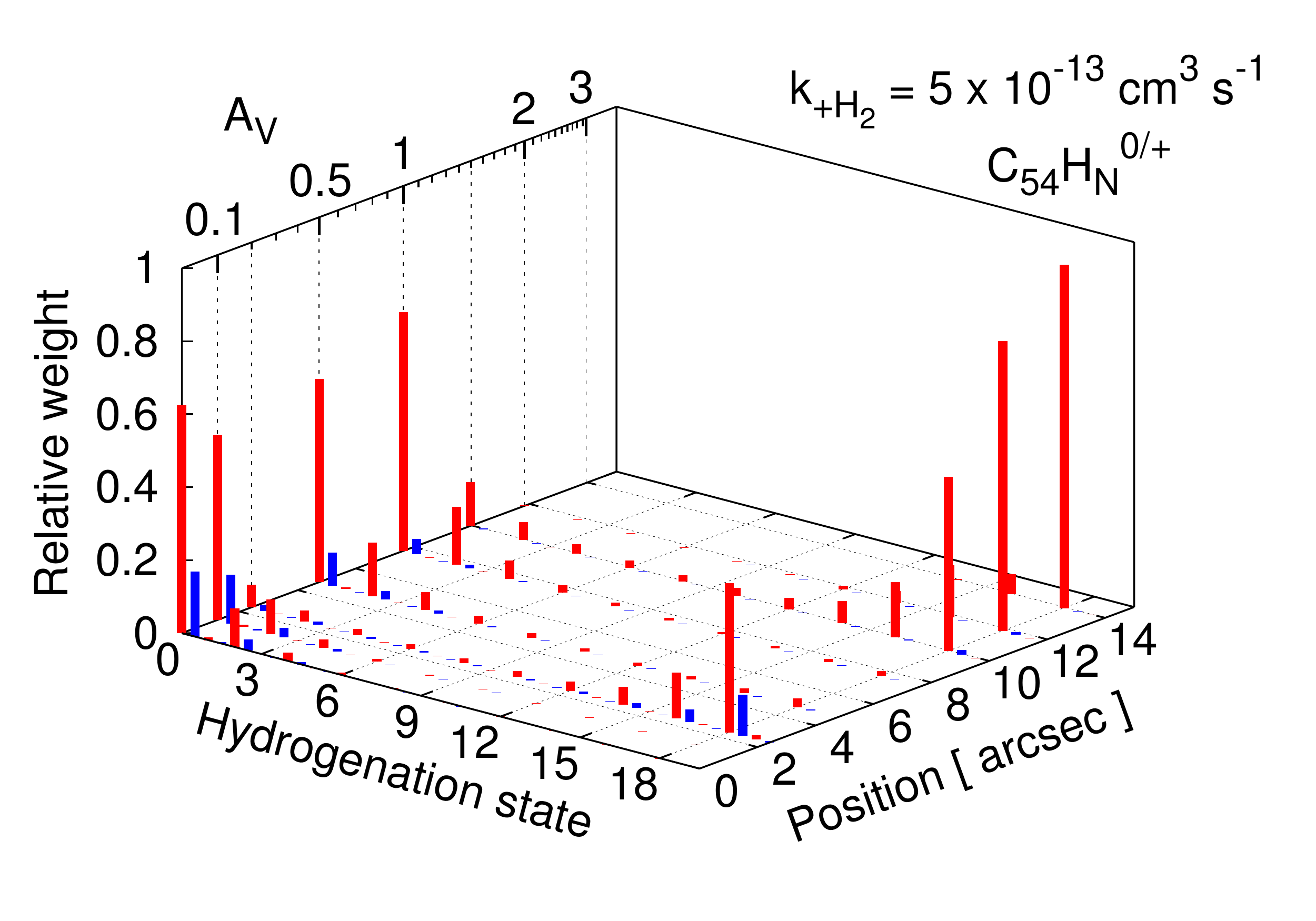} \hfill
         \includegraphics[width=0.49\textwidth, keepaspectratio]{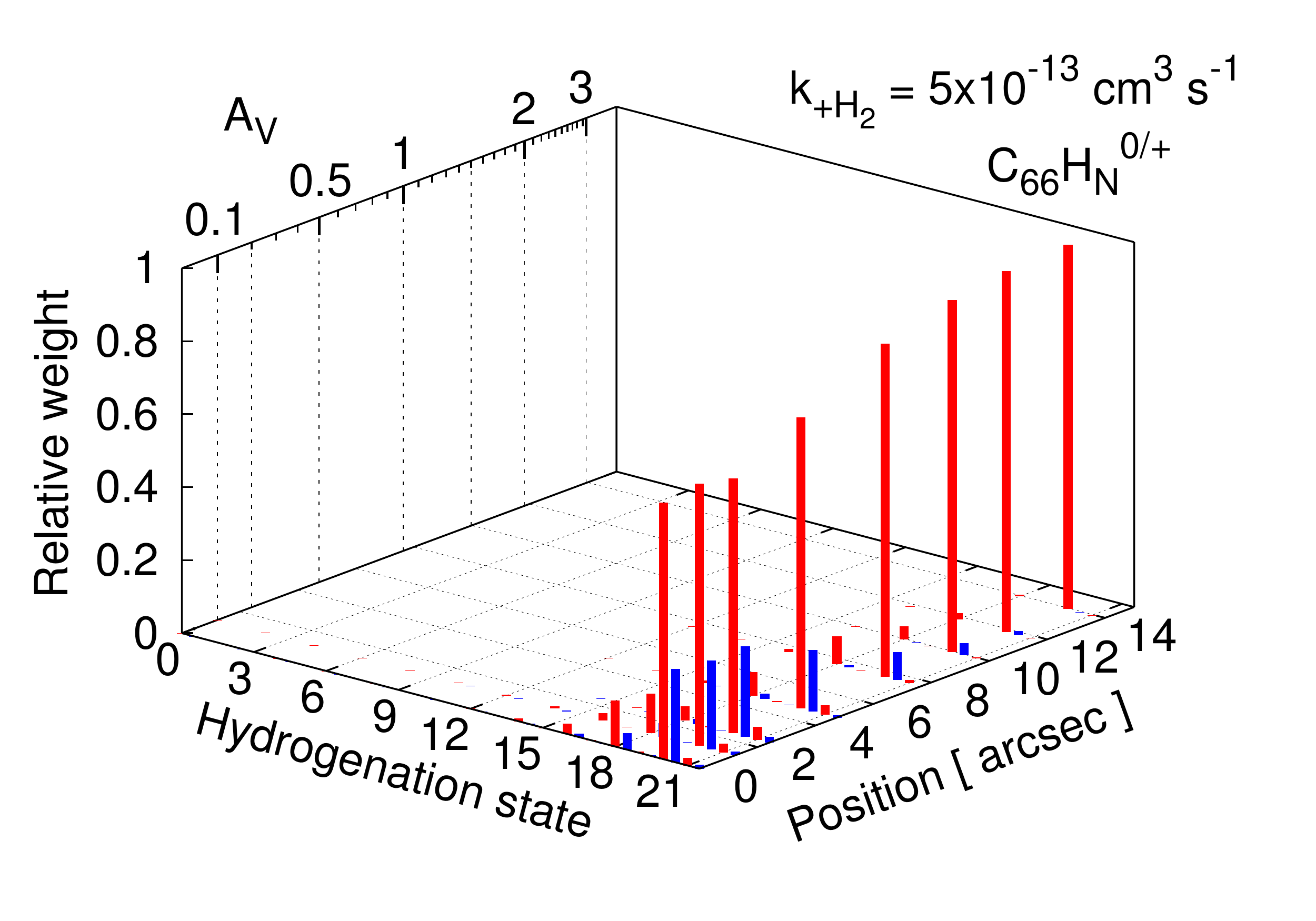} \\
         \includegraphics[width=0.49\textwidth, keepaspectratio]{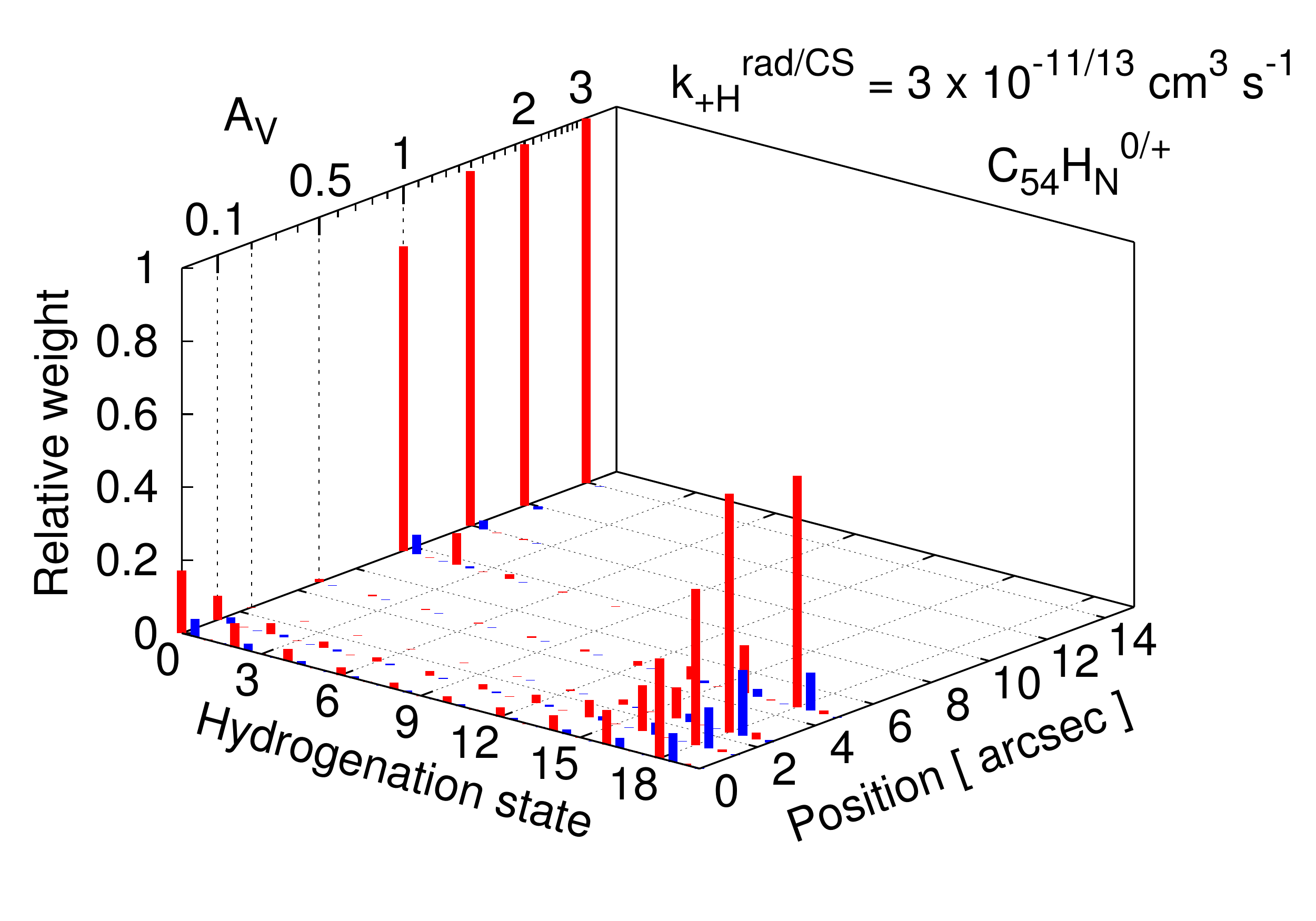} \hfill
         \includegraphics[width=0.49\textwidth, keepaspectratio]{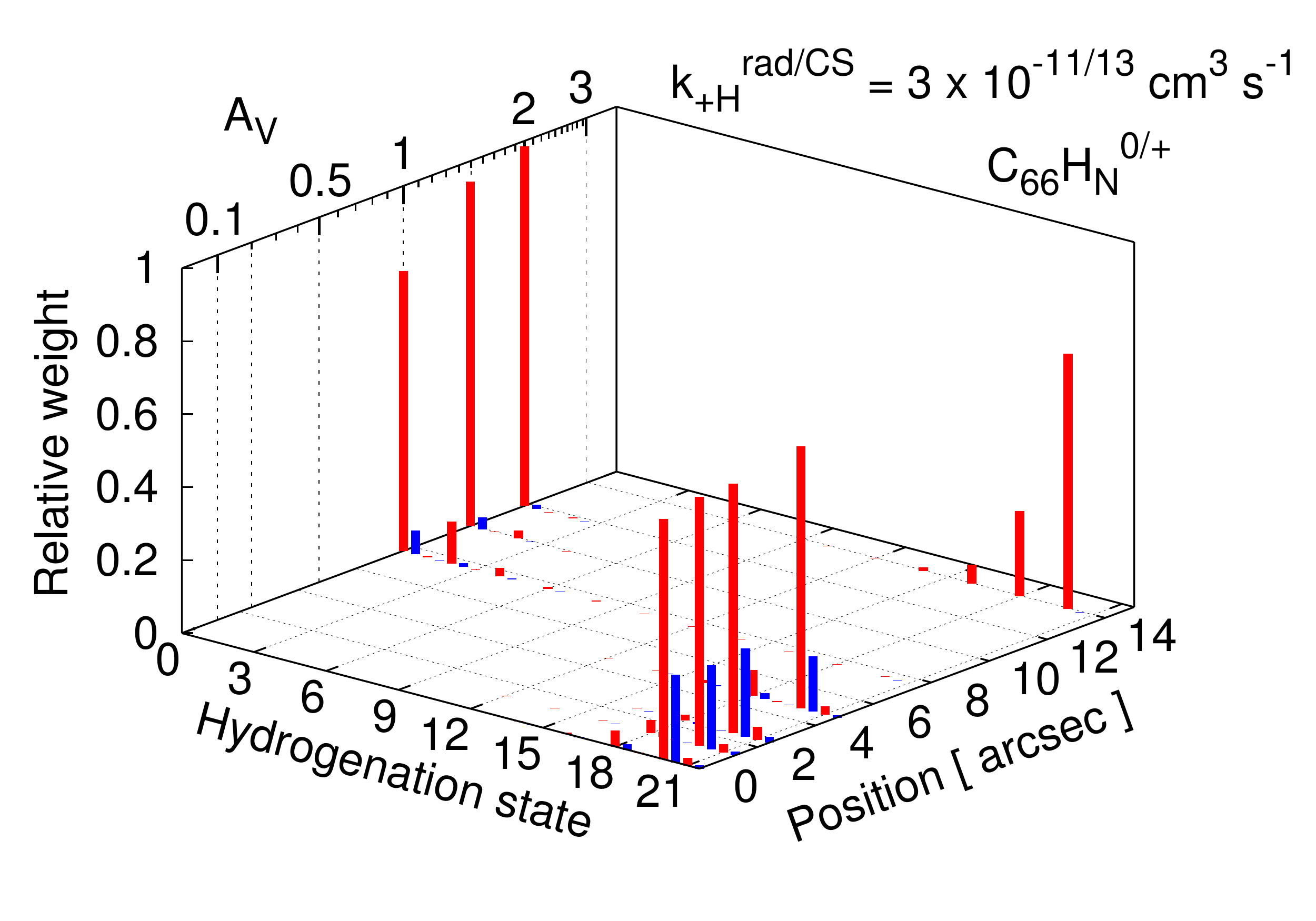}
      \end{center}
      \caption{Same as Fig.~\ref{fig:Results_standard} but with a reactivity of PAH cations with \hh of $k_{\rm +H_2}=5
      \times10^{-13}$\,cm$^3$\,s$^{-1}$ (upper panel), and with a reactivity of neutral PAHs with H of $k_{\rm +H}^{\rm rad}=3
      \times10^{-11}$\,cm$^3$\,s$^{-1}$ for radicals and $k_{\rm +H}^{\rm CS}=3 \times10^{-13}$\,cm$^3$\,s$^{-1}$ for 
      closed-shell (lower panel). Figures for coronene (\chp{24}{N}{+/0}) and circumcircumcoronene (\chp{96}{N}{+/0})
      are not shown as they exhibit only marginal variations with respect to Fig.~\ref{fig:Results_standard}. Neutral 
      species are in red, cations in blue.}
      \label{fig:Results_nonstandard}
   \end{figure*}



   The computed charge and hydrogenation states of coronene, circumcoronene, circumovalene and circumcircumcoronene 
   in the NGC 7023 NW PDR are shown in Figs.~\ref{fig:Results_standard} and ~\ref{fig:Results_nonstandard} 
   for different assumptions concerning the reactivity of PAHs with hydrogen. For all our calculations, we assumed 
   neutral normally hydrogenated PAHs as initial conditions. 
   In this section, we 
   present our results focusing successively on the charge state and the hydrogenation state of PAHs once steady state 
   is reached.

\subsection{Evolution of PAH charge in NGC 7023-NW}

\begin{figure}
      \begin{center}
         \includegraphics[width=0.49\textwidth, keepaspectratio]{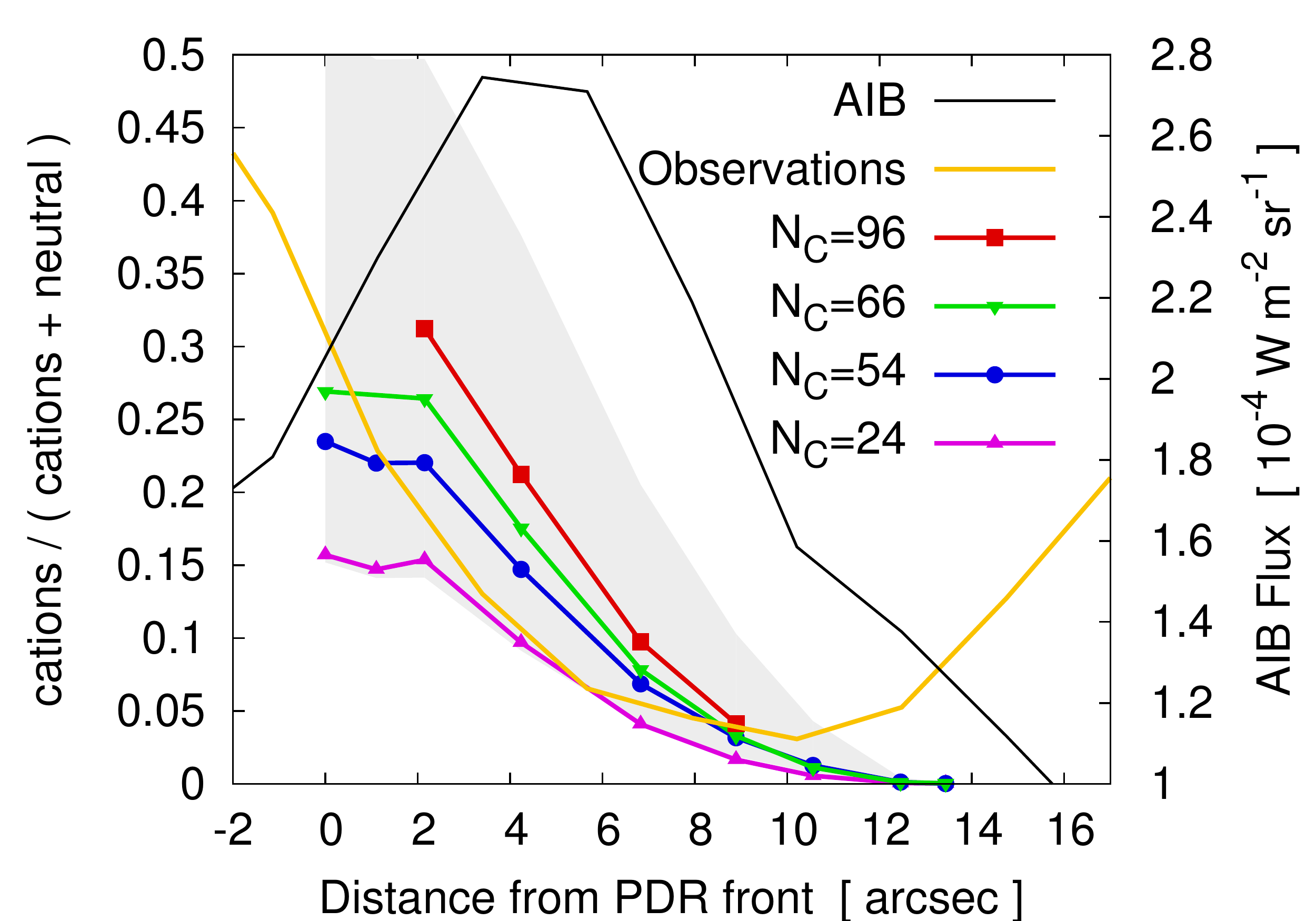}
      \end{center}
      \caption{Fraction of ionized species (PAHs and carbon clusters) along the Star-NW cut in the NW PDR as derived 
      from the observations \citep[yellow, from][]{pilleri_evaporating_2012} and as computed with our model (other 
      colours). The total AIB flux is also plotted in black. The grey surface shows the uncertainty range on the ionized 
      fraction for \chp{54}{N}{+} species (N=0-19), which reflects the range of values we considered for the electron
      recombination rate. The PDR front is assumed to lay at $42\arcsec$ from the star.}
      \label{fig:ChargeProfile}
   \end{figure}

   Using our model, we computed the global cation fraction for the four PAHs considered in this study by summing 
   the cation fraction in all hydrogenation states (e.g. for coronene $\sum_{N_{\rm H}}$ \chp{24}{N_{\rm H}}{+} / 
   $\sum_{N_{\rm H},q}$ \chp{24}{N_{\rm H}}{q} with $ N_{\rm H}=[0,12]$). In Fig.~\ref{fig:ChargeProfile}, the 
   calculated values are compared with the cation fraction extracted from the analysis of the AIB spectrum along the 
   Star-NW cut (Sect.~\ref{sec:APenvironment}).
    
   The computed cation fractions are in good agreement with the observed one in the cavity, as we predict fractions 
   above 90\% of cations at $20\arcsec$ from the star. In the region of the PDR where ionization of PAHs is found to be 
   significant (\Av  $\lesssim 2$), photoionization dominates the evolution of the charge state of PAHs since the UV 
   radiation field intensity  varies by a factor ${\sim10}$, whereas the PDR model predicts a rather constant abundance 
   of free electrons and a gas temperature that changes by a factor of 2 (see Fig.~\ref{fig:7023obs}), which results in 
   a variation of the rate of recombination of PAH cations with free electrons by a factor of less than 2. The agreement 
   between computed and observed cation fractions is also good within the AIB peak emission region (between $42\arcsec$ 
   and $52\arcsec$ from the star along the Star-NW cut), both in terms of absolute values and of profile evolution.
   Finally, the cation fraction is found to increase with the PAH size, $N_{\rm C}=96$ being twice more ionized than 
   $N_{\rm C}=24$. A more quantitative comparison  between modelling and observations would require to weight each 
   calculated charge ratios by the column density of the corresponding PAH along the line of sight.

\subsection{Hydrogenation state}

   As shown in Fig.~\ref{fig:Results_standard}, the hydrogenation state of PAHs evolves significantly with both PAH size 
   and position within the PDR. A general result is that partially dehydrogenated species never dominate in terms of 
   abundances. Even collectively, their maximum weight is 40\%, and for a very narrow range of spatial positions. 
   Therefore, the resulting hydrogenation states can be classified into four categories: pure carbon clusters 
   ($N_{\rm H}=0$), partially dehydrogenated PAHs ($1 \leq N_{\rm H} \leq N_{\rm H}^0 - 1$), normally hydrogenated PAHs, 
   and superhydrogenated PAHs ($N_{\rm H}=N_{\rm H}^0$+1). This behaviour results from a general trend that more 
   dehydrogenated PAHs are less photostable. Therefore, once a normally hydrogenated PAH starts losing H-atoms, further 
   dehydrogenation is even faster. 
   
   \begin{figure}
         \includegraphics[width=0.49\textwidth,  keepaspectratio]{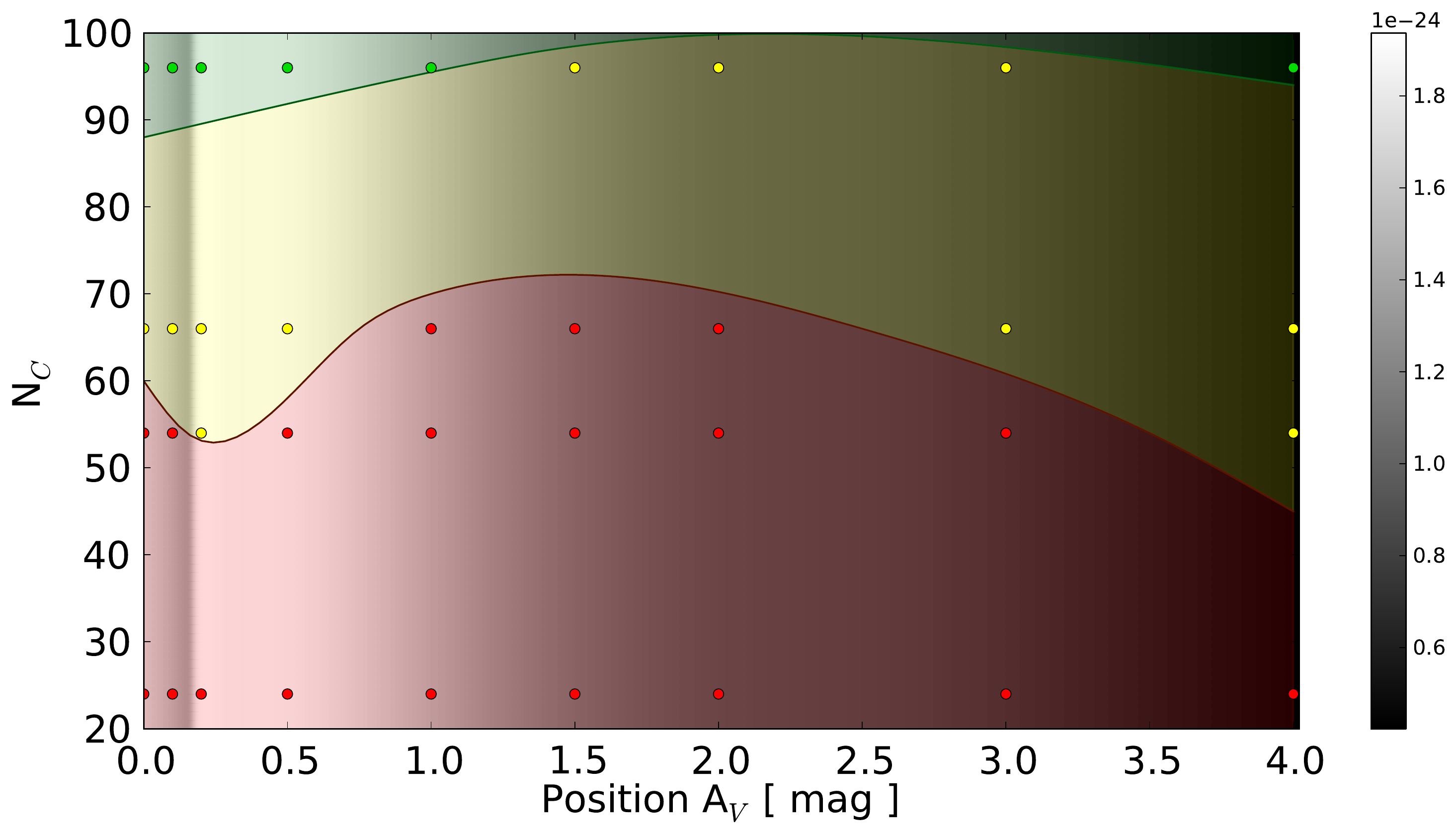}
         \includegraphics[width=0.49\textwidth,  keepaspectratio]{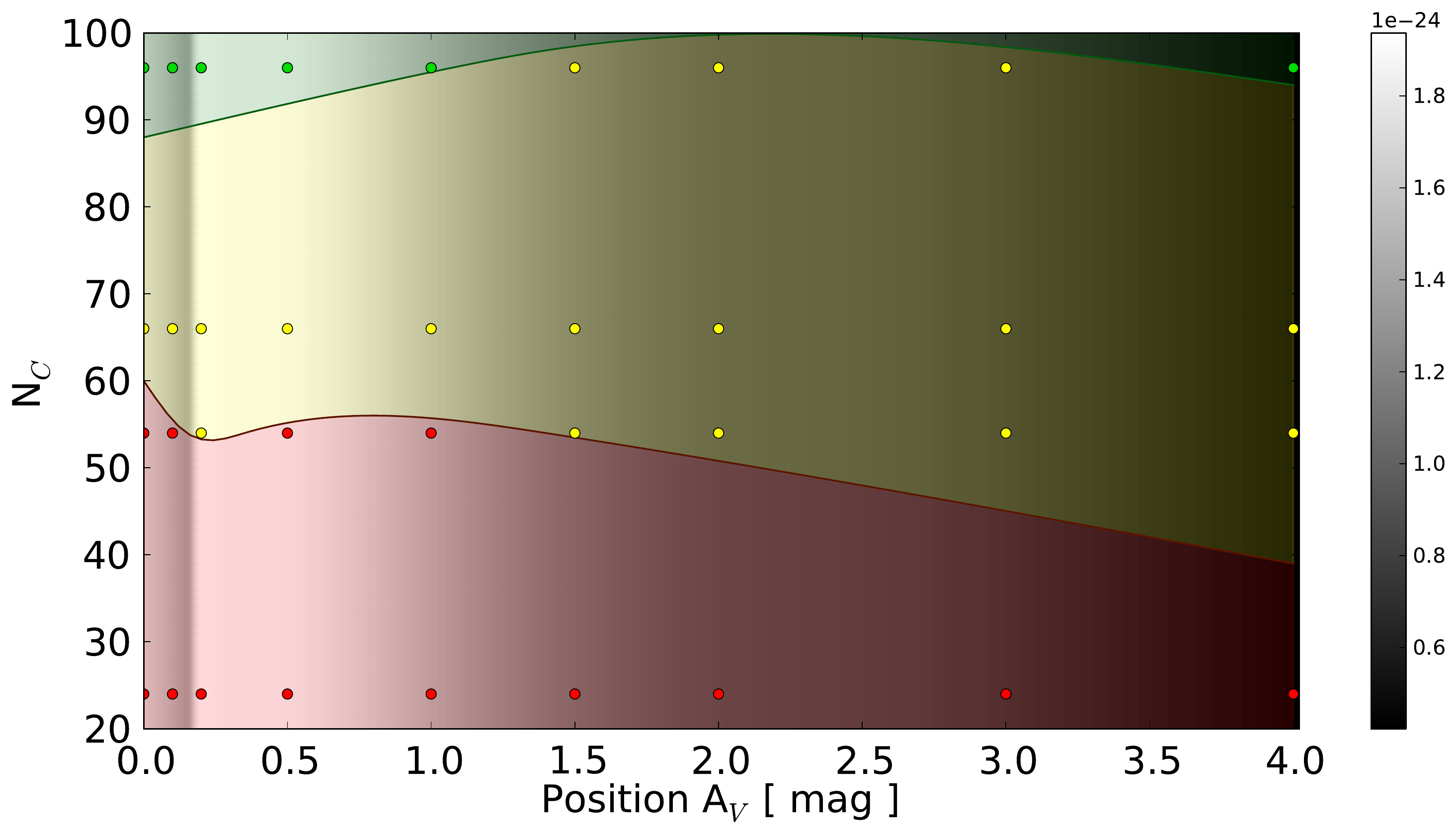}
         \includegraphics[width=0.445\textwidth, keepaspectratio]{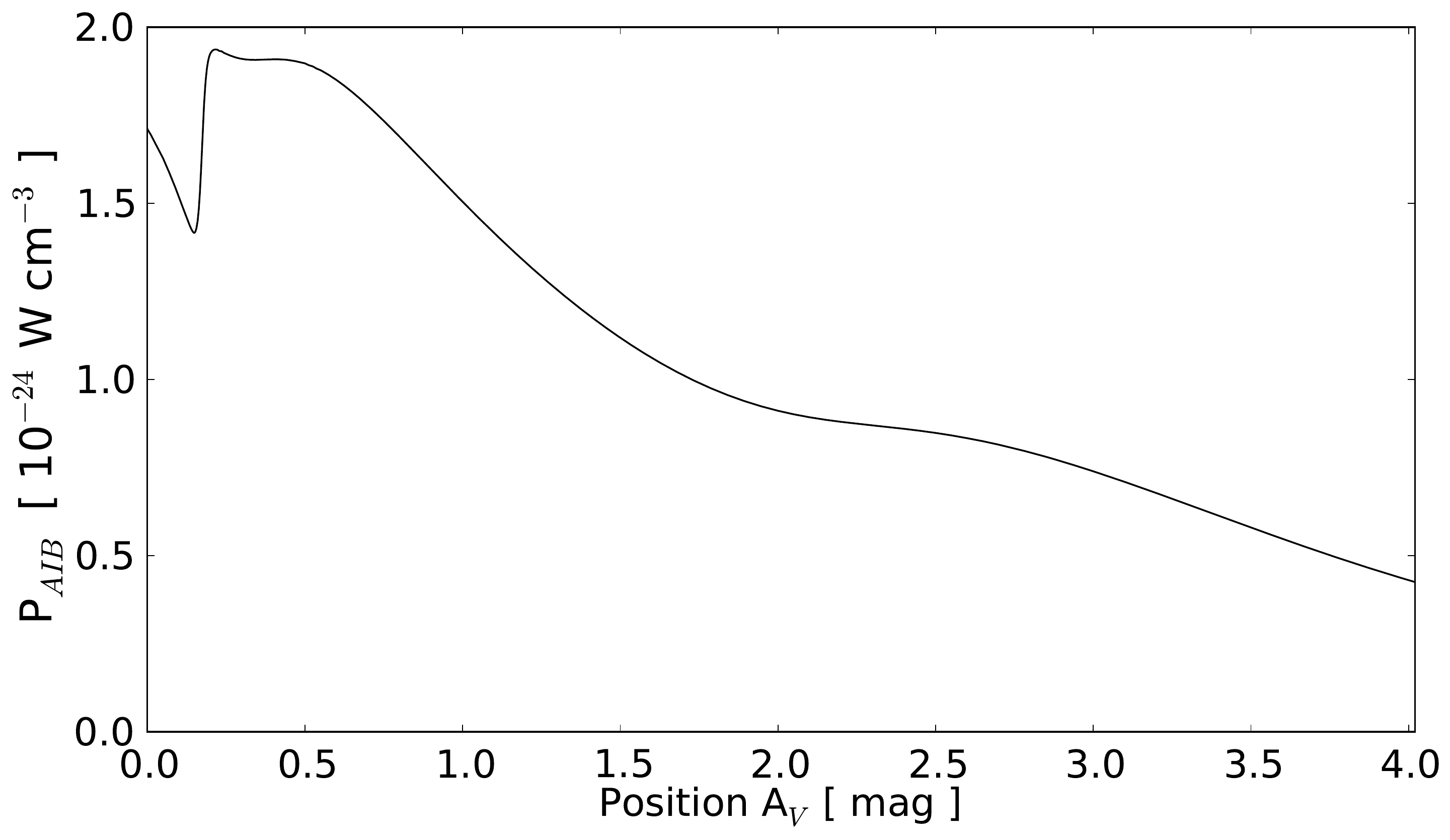}
      \caption{Dominant hydrogenation state as a function of PAH size and position in the NGC\,7023 NW PDR, without
      reactivity (\textit{upper panel}) or with a rate of $5\times10^{-13}$\,cm$^3$\,s$^{-1}$ (\textit{middle panel}) 
      for the reaction between PAH cations and \hhp. The points represent the parameters for which models were computed. 
      Red, yellow and green represent carbon clusters, normally hydrogenated PAHs, and superhydrogenated PAHs (limited
      to one extra H-atom), respectively. 
      Lines delimiting the different domains were interpolated to guide the eye and should be considered 
      with caution. The power per unit of volume $P_{\rm AIB}$ emitted in the mid-IR domain by PAHs and carbon 
      clusters as a function of their position is shown in the {\it lower panel}, and is also overlaid as a brightness 
      scale to the upper and middle panels. The variation observed between \Av=0 and 0.2 reflects the variation of $\nH$
      (see Fig.~\ref{fig:7023obs}).}
      \label{fig:Evol_Taille}
   \end{figure}

   We summarize the evolution of the dominant hydrogenation state as a function of the position and size of the species 
   in Fig.~\ref{fig:Evol_Taille} (results in the cavity are not shown). To illustrate the contribution of each population
   to the AIB emission, Fig.~\ref{fig:Evol_Taille} also shows $P_{\rm AIB}$, the emission power per unit volume in the 
   mid-IR domain.
   It was calculated using the formula $P_{\rm AIB} = \epsilon \, \nH \, G_0$ \citep{pilleri_evaporating_2012}, where
   $\epsilon = 5 \times 10^{-32} \, {\rm W\,H}^{-1}$ and $G_0$ is expressed in units of Habing. Note that in this 
   formula, all species, whether PAHs or carbon clusters, are assumed to emit the same power per C-atom. For a given 
   position in the PDR, larger molecules tend to be more hydrogenated than smaller, as expected. For a given size, 
   however, one find alternatively hydrogenated and dehydrogenated species when increasing the distance from the star, 
   whereas a monotonous behaviour, as a function of the intensity of the UV radiation field, could be expected. 
   Similarly, superhydrogenated PAHs dominate close to the cloud surface, and are less abundant in deeper layers. These 
   inversion features are due to the decrease in atomic hydrogen density from \Av=0.2 to \Av$\sim2$, and is therefore a 
   consequence of both the environment evolution and the assumption that reactivity with \hh can be neglected. We 
   discuss this assumption further in Sect.~\ref{sec:sensib_process}.
   
   Superhydrogenated PAHs are found to be abundant only for the largest species ($N_{\rm C}=96$), despite the fact that
   we considered a lower limit for the dissociation rate of superhydrogenated PAHs. Pure carbon clusters are found to be
   the dominant form for small- ($N_{\rm C}<50$) and, possibly, medium-sized ($50 \lesssim N_{\rm C}\lesssim 70$) species, 
   until cloud depths, at which PAHs are thought to be part of very small grains \citep[\Av$\gtrsim$2, see Fig.~\ref
   {fig:7023obs} and][]{pilleri_evaporating_2012}. In the cavity, where the density is lower and the radiation field 
   much stronger, we predict all PAHs to be completely dehydrogenated.
   
\begin{figure}
      \begin{center}
         \includegraphics[width=0.5\textwidth, keepaspectratio]{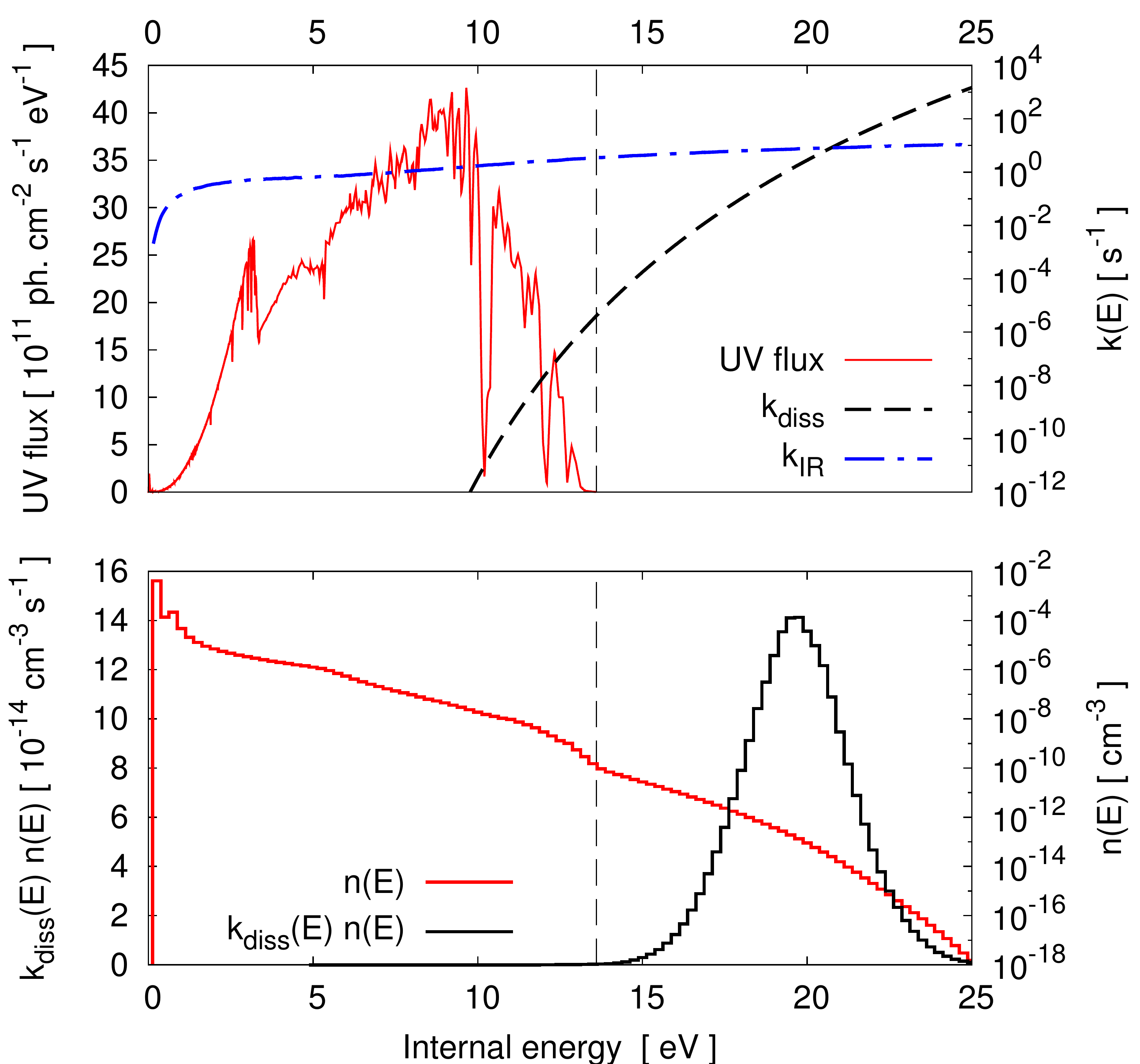}
      \end{center}
      \caption{\textit{Upper panel:} Spectrum of the local UV-visible radiation field in NGC 7023-NW at $42\arcsec$
      (\Av=0) from the star, compared to the rates of dissociation and IR emission of the circumovalene molecule 
      \ch{66}{20}.
      \textit{Lower panel:} The internal energy of \ch{66}{20} (red histogram) computed with our model in 
      steady state, in the radiation field showed in the upper panel. The black histogram shows the 
      distribution of internal energies of \ch{66}{20} molecules prior to dissociation. The vertical dashed lines show
      the 13.6 eV limit, emphasising that dissociation is dominated by multiphoton events.
      }
      \label{fig:Ediss}
   \end{figure}

   Interestingly, circumcoronene and circumovalene are found to be fully dehydrogenated in the NW PDR at \Av=0, and 
   between \Av=1 and \Av=2, respectively, whereas for the same conditions previous models \citep{le_page_hydrogenation_2001, 
   berne_formation_2012} would predict normally hydrogenated circumovalene to dominate. This discrepancy is due to 
   multiphoton events that are taken into account in our model. Multiphoton events are by far the main channel for PAH 
   dissociation at threshold energy (where IR photon emission and dissociation are equally probable) above the Lyman 
   limit. Figure~\ref{fig:Ediss} shows in the case of circumovalene \ch{66}{20} that cooling by IR photon emission is 10 
   orders of magnitude faster than dissociation at wavelengths of maximum UV radiation. \ch{66}{20} molecules bearing an 
   internal energy close to the threshold energy ($\sim20$ eV) after the successive absorption of several UV photons are 
   found to be the most numerous to dissociate, despite their very low abundances (cf. Figure~\ref{fig:Ediss}).

\subsection{Timescales}\label{sec:timescales}
   
   The results presented in the previous sections are obtained after steady state has been reached. We investigate here 
   how steady state timescales compare with other evolutionary timescales in a given environment. The steady state for 
   PAH charge evolution is reached within a few hours at the position \Av=1. As a comparison, for these conditions, the 
   typical timescale for C photoionization is $\sim$1 month \citep{le_petit_model_2006}. This timescale can also be 
   compared with the timescale for charge exchange (c.e.) between neutral PAHs and C$^+$. In the conditions of NGC 7023 
   NW, using $k_{\rm c.e.} = 3\times10^{-9}$ cm$^{3}$\,s$^{-1}$ from \citet{canosa_reaction_1995}, $\nH = 10^4$ cm$^{3}$, 
   C/H = $1.3 \times 10^{-4}$ from our PDR calculations, and assuming that all atomic carbon is ionized, we find $\tau = 
   1/(k_{\rm c.e.}[{\rm C}^+]) \sim 8$ years. This shows that charge exchange is negligible in these conditions.
   
\begin{figure*}
      \begin{center}
         \includegraphics[width=0.49\textwidth, keepaspectratio]{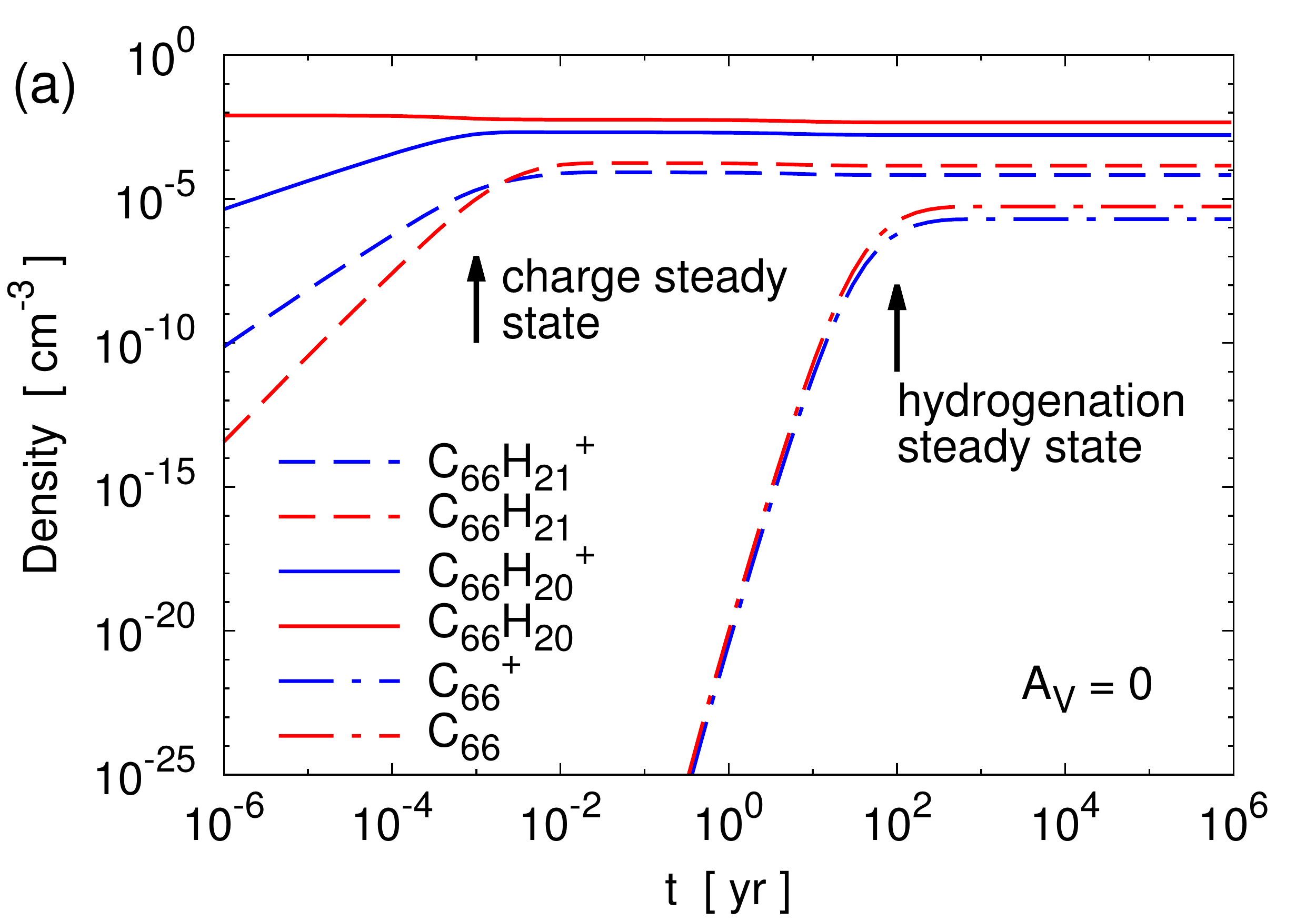}
         \includegraphics[width=0.49\textwidth, keepaspectratio]{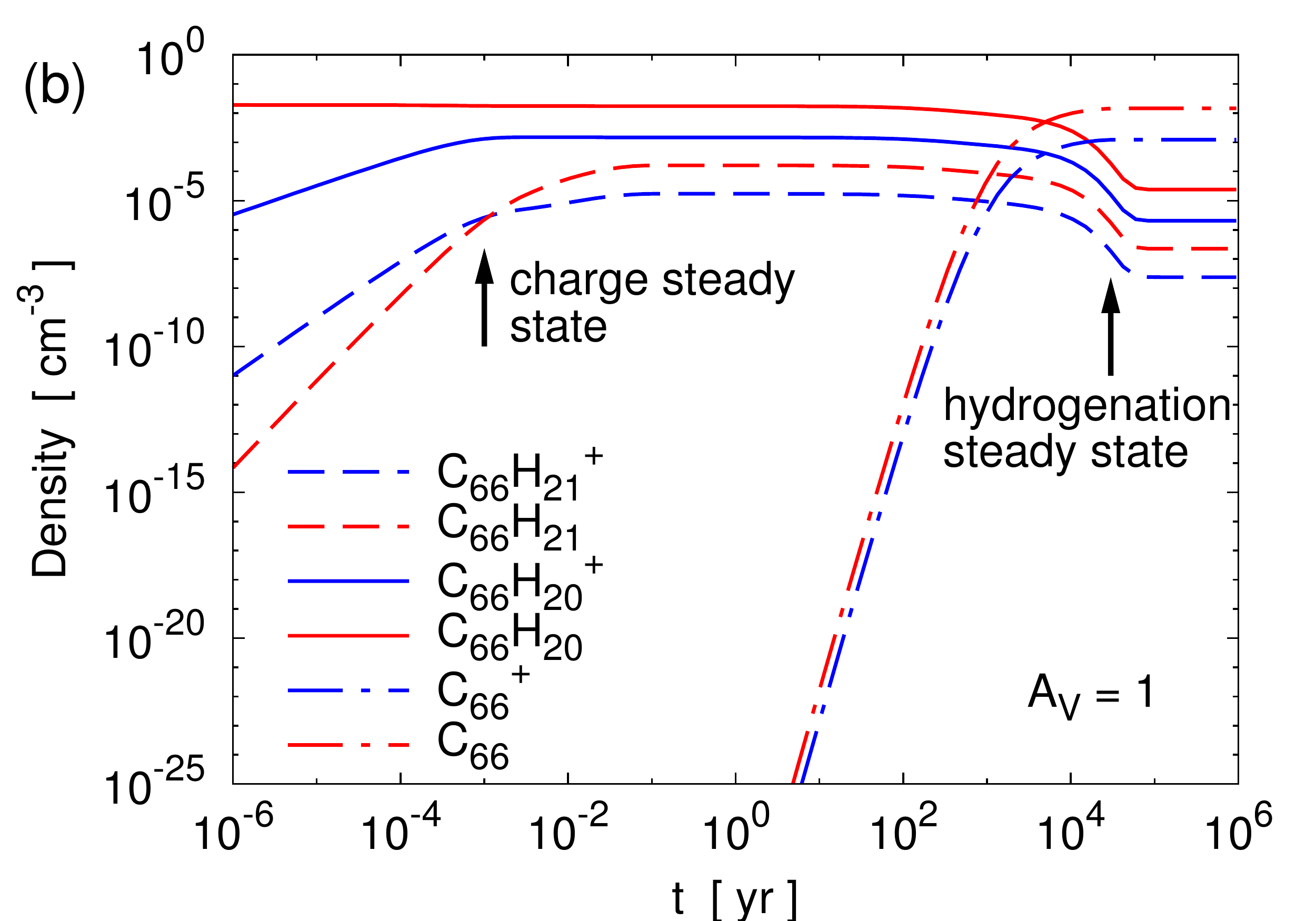}
      \end{center}
      \caption{Time evolution of some charge and hydrogenation states of circumovalene in the NGCC\,7023 NW PDR, at depths
      corresponding to \Av = 0 (a) and \Av = 1 (b), i.e. $\sim42$ and $\sim50 \arcsec$ from the star, respectively. The typical 
      timescale to reach steady state for charge and hydrogenation states are indicated with black arrows.}
      \label{fig:timescale}
   \end{figure*}

   The steady state for PAH hydrogenation is reached for all species on much longer timescales, which span a very broad 
   range of values depending on PAH size. Small species reach there steady state within less than a year (e.g. $\sim$1 
   month for coronene at \Av=1), while large species exhibit very long timescales, as can be seen in figure~\ref
   {fig:timescale}, with $\sim10^2$ years at \Av=0 and $\sim10^4$ years at \Av=1 in the case of circumovalene, if no 
   reactivity is considered with \hhp. When considering the reactivity of PAH cations with \hh at a rate of $k_{\rm +H_2}=5
   \times10^{-13}$\,cm$^3$\,s$^{-1}$, the timescale at \Av = 1 decreases down to $\sim10^2 - 10^3$ years. These values 
   are very large with respect to the typical hydrogenation time (1/($k_{+\rm H} \times \nH) \sim 10$ days), and seem 
   surprising at first sight. In fact, this is linked to the absence of dominating partially hydrogenated species: 
   the global time evolution for one PAH species scales with the typical times needed for it to evolve through 
   \textit{all} its partially hydrogenated states, from the normally hydrogenated state to the fully dehydrogenated 
   state and \textit{vice versa}. 
   These timescales become long when multiphoton events are required to dissociate PAHs, as each dehydrogenation step
   behaves as a limiting step. 
   This illustrates that the complex evolution of the hydrogenation state of PAHs requires a detailed modelling.
   
   Considering that \citet{alecian_characterization_2008} estimated the age of HD 200775 around 10$^5$ years, our 
   results suggest that the evolution of large PAHs is coupled to the dynamics of the region, and that large PAHs may 
   never reach their steady state abundances. 
      
\section{Sensitivity to parameters}\label{sec:sensitivity}

\subsection{Identifying the key processes}\label{sec:sensib_process}
   
   We have discussed in Sect.~\ref{sec:PCprocess} the uncertainties on the parameters of the various processes. In this 
   section, we aim at evaluating the sensitivity of the results to some of them, in order to identify the processes that
   require further investigation.
   
   The determination of the charge state of PAHs is mainly limited by the knowledge of the rate $\alpha$ for their 
   recombination with electrons.  In the case of circumcoronene, we computed additional models with the upper and lower
   values of $\alpha$ as defined in Sect.~\ref{recomb}. In figure~\ref{fig:ChargeProfile}, we represented by a grey area 
   the domain of values resulting from this uncertainty. Thus, the fraction of PAH cations is determined with a typical 
   uncertainty of a factor of two, which is slightly larger than the dispersion of cation fraction between PAHs of 
   different sizes. This means that further constraints on the PAH size distribution using PAH charge states would 
   require more investigation of the electronic recombination process.

   The results presented in Fig.~\ref{fig:Results_standard} do not consider any reactivity of PAHs with \hhp, and of
   neutral PAHs with H. Figure~\ref{fig:Results_nonstandard} shows the influence of these two processes. The reactivity 
   of neutral PAHs impacts the first layers of PDRs, at the interface with the cavity where atomic hydrogen 
   dominates, whereas the reactivity of PAH cations with \hh plays a significant role deeper in the PDR, between \Av=0.5 
   and 3. The smallest in size species ($N_{\rm C}=24$) are not affected by these processes as they remain fully 
   dehydrogenated. Similarly, the high photostability of the largest species ($N_{\rm C}=96$) ensures them to remain 
   normally hydrogenated regardless of these additional processes. On the contrary, intermediate species can be
   significantly more hydrogenated. The size - position diagrams (Fig.~\ref{fig:Evol_Taille}) shows that the uncertainty
   on the reaction rates with hydrogen leads to some uncertainty on the position of the transition between normally 
   hydrogenated PAHs and carbon clusters, which lays between $N_{\rm C}\sim50$ and $\sim60$ at \Av=0 and between 
   $N_{\rm C}\sim50$ and $\sim70$ at \Av=1. Therefore, the lack of knowledge on these processes, and particularly the 
   reactivity with \hhp, appears as a major limitation for predicting the minimum size of PAHs in PDRs.


\subsection{Sensitivity to the astrophysical conditions}\label{sec:sensib_astro}

   Deriving the general trends of PAH evolution as a function of the astrophysical conditions from the case study of
   NGC 7023-NW is not straightforward, because the evolution of the physical conditions are correlated 
   along the Star-NW cut. Therefore, in addition to the case study of NGC 7023, we also followed the method proposed by 
   \citet{le_page_hydrogenation_2001} and computed the charge and hydrogenated states of the four studied PAHs on a grid 
   of astrophysical conditions. We used the interstellar radiation field (ISRF) of \citet{mathis_interstellar_1983} as 
   UV-visible spectrum, and normalised it to integrated intensities of $10^n$ ($n=0,1,2,3,4$) in Habing units. 
   We considered a gas temperature of 100 K, and several values of $\nH$ between 10 and 10$^5$ cm$^{-3}$, with constant 
   fractions of atomic hydrogen $n({\rm H})=0.5 \nH$ and free electrons $n({\rm e}^-)=1.4\times
   10^{-4} \nH$. 

%

\begin{figure}
      \begin{center}
         \includegraphics[width=0.5\textwidth, keepaspectratio]{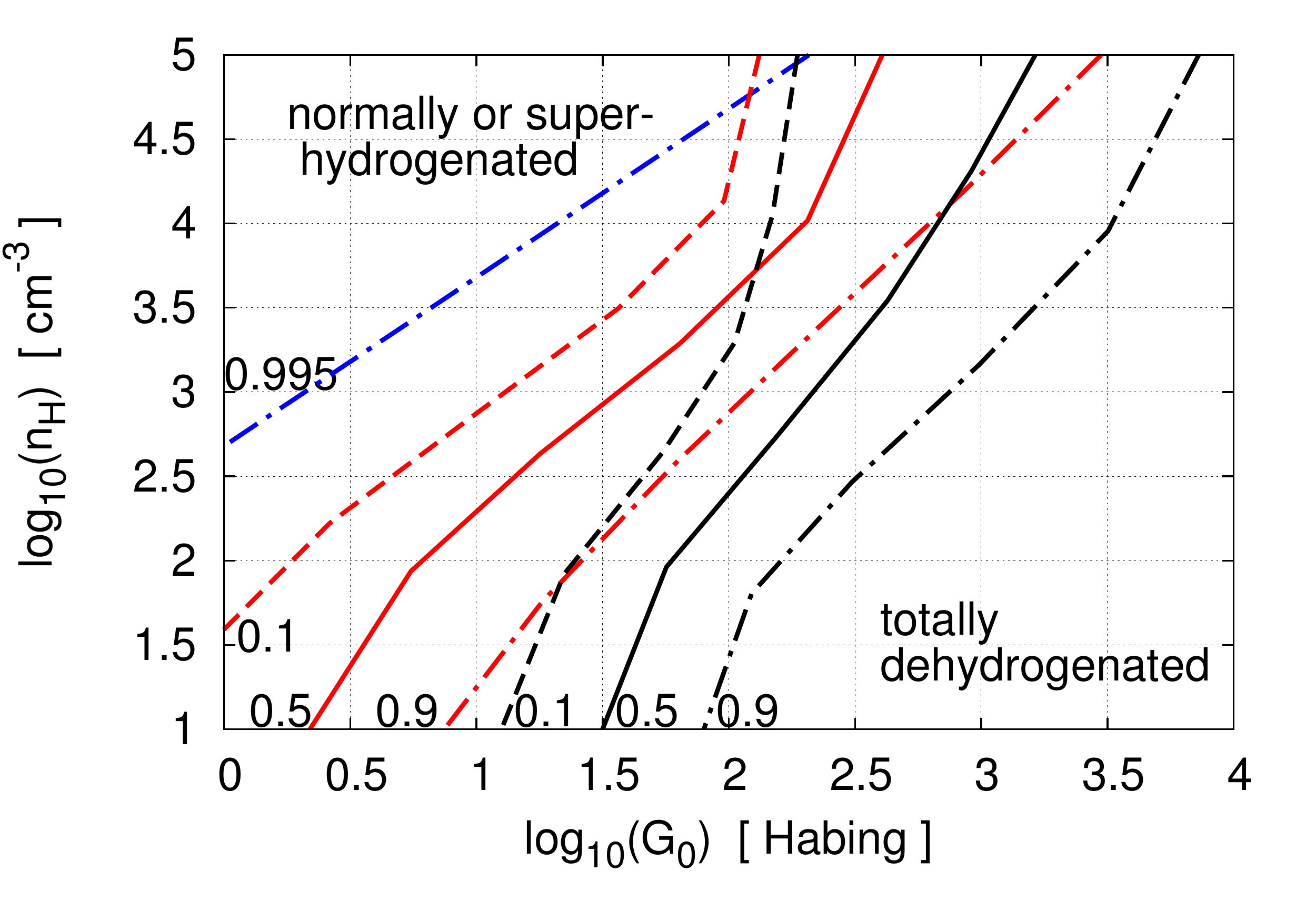}
      \end{center}
      \caption{Hydrogenation state of circumcoronene (\ch{54}{18}, red) and circumovalene (\ch{66}{20}, black) as a 
      function of the intensity of the UV radiation field $G_0$ in Habing units and the local density $\nH$. The lines
      labelled with 0.1 (respectively 0.5, 0.9) show the points for which 10\% (respectively 50\%, 90\%) of the species 
      are totally dehydrogenated. For coronene, only the blue line labelled with 0.995 where 99.5\% of coronene is 
      fully dehydrogenated is represented, which is the dominant state of coronene for all the conditions spanned here.}
      \label{fig:critical_G0_nH}
   \end{figure}

   The results are presented in Fig.~\ref{fig:critical_G0_nH}. They show that the evolution of the hydrogenation state 
   of \ch{54}{18} and \ch{66}{20} exhibits a complex behaviour with $G_0$ and $\nH$. On average, the slope of the 
   boundary between the hydrogenated and dehydrogenated domains (solid line in Fig.~\ref{fig:critical_G0_nH}) is clearly 
   greater that 1 ($\sim$1.5 and $\sim$2 for circumcoronene and circumovalene, respectively) in a log-log diagram. This
   results from the fact that, on average, roughly two photons are needed for \ch{54}{18} and \ch{66}{20} to reach their 
   threshold energies for fragmentation at $\sim17$ and $\sim20$\,eV, respectively. When the intensity of the UV radiation 
   field increases, the proportion of multiple absorption events increases and the probability of low energy photons being
   involved increases, which leads to slope changes with $G_0$ in Fig.~\ref{fig:critical_G0_nH}.
   
   Coronene is the only species for which we find that the hydrogenation state scales with $G_0/\nH$, similarly to the 
   results reported by \citet{le_page_hydrogenation_2003}, who did not include multi-photon events in their model. This 
   is due to the fact that coronene has a small threshold energy for fragmentation ($E_{\rm th}\approx9.6<13.6$ eV) and 
   therefore can easily dissociate after the absorption of a single photon. In other words, $G_0/\nH$ is a correct 
   parameter to quantify the hydrogenation state of PAHs only in cases in which dissociation is achieved with a single 
   UV photon. However, in such cases dehydrogenation is so fast that it leads to the complete dehydrogenated product. 
   Therefore, the population of interstellar PAHs must be dominated by species for which multi-photon events dominate 
   the photodissociation process, and whose hydrogenation state is not well parametrised by $G_0/\nH$.

\section{Discussion and perspectives}\label{sec:discussion}



\subsection{Consequences for the identification of a specific PAH in the ISM}

   The detection of specific PAHs in the ISM has remained unsuccessful: \citet{pilleri_search_2009} reported an upper 
   limit for rotational lines of corannulene \ch{20}{10} in the Red Rectangle (RR) nebula, and \citet{kokkin_optical_2008} 
   also reported an upper limit on the abundance for hexa-peri-hexabenzocoronene \ch{42}{18} in the diffuse ISM from its
   optical emission. As a comparison, the results presented in Fig.~\ref{fig:critical_G0_nH} show that the larger,
   more compact, and therefore more photostable circumcoronene molecule \ch{54}{18} is expected to be at the edge between
   normally hydrogenated and fully dehydrogenated in physical conditions typical of the diffuse ISM (e.g. $G_0 \sim$ 1 
   and $\nH\sim$10-100 cm$^{-3}$). Thus it is very likely that \ch{42}{18} is completely dehydrogenated in
   the diffuse ISM and this could explain why it was not detected by \citet{kokkin_optical_2008}. 
   
   In the particular case of \ch{20}{10} in the RR nebula, the situation is less clear as the radiation field of this 
   object is remarkably poor in UV photons, so that Fig.~\ref{fig:critical_G0_nH} cannot be used for this object. Using 
   the radiation field spectrum proposed by \citet{mulas_general_2006}, a density of $\nH=10^{5}$ cm$^{-3}$ with an 
   arbitrary fraction of molecular hydrogen of 0.5, and a temperature of 100 K \citep{bujarrabal_orbiting_2005}, our 
   model predicts coronene \ch{24}{12} to be completely dehydrogenated in RR. It is therefore very likely that 
   corannulene \ch{20}{10} cannot survive in RR as well. As a consequence, our results strongly suggest that future 
   search for PAH spectroscopic signatures should focus on much larger species, containing at least 50 or 60 carbon 
   atoms.

\subsection{The carbon content in PAHs and carbon clusters}

   The fact that we predict a large abundance of carbon clusters in the cavity and in a significant fraction of the PDR 
   of NGC 7023 NW raises the question of the evolution of carbon content in PAHs and carbon clusters. Carbon clusters 
   have properties that differ significantly from those of PAHs, in terms of structure, stability and spectroscopy. 
   Therefore, when a PAH is photodissociated until full dehydrogenation, it can be considered as a loss of carbon 
   content in the PAH population and a gain for the carbon cluster population. However, the latter population also may 
   undergo evolution via photodissociation and carbon accretion. To get a first estimate, we used  our model for 
   computing the effective photodissociation rates (number of dissociation events per time unit) of planar C$_{24}^{0/+}$, 
   C$_{66}^{0/+}$ and C$_{96}^{0/+}$ in the cavity and the NW PDR. We used the IR photon emission rates presented in 
   Sect.~\ref{kIR} and computed the dissociation rates for the losses of C, C$_2$ and C$_3$ using Eq.~\ref{eq:kdiss} and 
   the parameters proposed by \citet{leger_photo-thermo-dissociation._1989} that are based on thermodynamic properties 
   of graphite.
   
   




   We found that the largest cluster in this study, C$_{96}^{0/+}$, is destroyed within a time longer than the age of 
   the star, even at $10\arcsec$ from the star. C$_{66}^{0/+}$ needs more than several 10$^6$ years to be significantly 
   photodissociated in the NW PDR, and its destruction rate rises steeply in the cavity until very fast photodestruction 
   ($\sim10^6$s$^{-1}$) as multiphoton events become frequent for the corresponding radiation field. C$_{24}^{0/+}$ 
   photodissociates much faster than larger species in the NW PDR.  Its photodissociation rate increases less rapidly than
   for larger species when going closer to the star, and reaches the same value as for C$_{66}^{0/+}$ close to the star. 
   Here again, the evolution of the photodestruction rate is highly non-linear due to the predominance of multiphoton 
   events (threshold energies for fragmentation are $\sim$15, 37, 54 eV for C$_{24}^{0/+}$, C$_{66}^{0/+}$ and 
   C$_{96}^{0/+}$, respectively). 

   These results draw a picture in which the cavity is populated by large carbon clusters while the PDR hosts a blend of
   large PAHs and relatively small carbon clusters. Carbon clusters are expected to explore various isomeric forms 
   including non planar (cage) ones \citep[see for instance ][for the various forms of C$_{24}$]{jones_structure_1997}.
   Heating by UV photons favours the exploration of the various isomeric forms. Our results are therefore consistent 
   with the formation of the cage fullerene C$_{60}$ in the cavity following the destruction of PAHs as suggested 
   by \citet{berne_formation_2012}.
   
   In addition, this raises the question of the contribution of carbon clusters to the AIBs, as well as to the physical 
   evolution of PDRs, e.g. through the photoelectric effect. Further consideration on these questions calls for more
   laboratory studies on the photophysical and chemical properties of these species.

\subsection{Perspectives}\label{sec:other_persp}

   In this paper, we have focused on the interaction with photons and the reactivity with electrons and hydrogen.
   Including a larger set of chemical processes would be of interest for future studies. 
   
   The reactivity of PAHs with C$^+$ has been studied experimentally by \citet{canosa_reaction_1995}. The authors found 
   the reaction to proceed at a rate close to the rate of Langevin with, in the case of anthracene, two possible channels 
   of comparable branching ratio: charge exchange (c.e.) and carbon accretion. In Sect.~\ref{sec:timescales}, we have 
   evaluated the timescale for c.e. to be $\sim8$ years in the conditions of NGC 7023 NW. A similar timescale is expected 
   for carbon accretion. These values are shorter than the timescale to reach hydrogenation steady state, which indicate 
   that the corresponding reactions have to be included in a PAH chemical model. Still, reactions of PAHs with C$^+$ have 
   been studied only for small PAHs and significant variations with the size of PAHs are observed in the branching ratio 
   between c.e. and carbon accretion, calling for investigation of the reactivity of larger PAHs, including 
   dehydrogenated PAHs, with C$^+$.
   
   PAHs can also react with atomic oxygen but these reactions are also poorly known. \cite{le_page_gas_1999} considered 
   a reaction rate for PAH cations  of $k_{\rm +O} \sim 10^{-10}$ cm$^{3}$\,s$^{-1}$ for the addition of an O-atom,
   regardless of the size of PAH cations. Assuming a density $\nH=10^4$cm$^{-3}$ and an O/H ratio of $3 \times 10^{-4}$, 
   this leads to a reaction timescale of $\sim 100$ years. Therefore, for large PAHs, reaction with oxygen could play a 
   significant role in PAH evolution.

   Finally, it might be interesting to investigate the possibility that PAHs form clusters and/or complexes with heavy atoms
   such as iron and silicon as proposed by several authors
   \citep{rapacioli_formation_2006, simon_photodissociation_2009, joalland_signature_2009}

\section{Conclusion}\label{sec:conclusion}

   In this paper, we presented a new model dedicated to the evolution of the hydrogenation and charge states of PAHs.
   This model
   was designed to compute the time evolution of the internal energy of species, so that multi-photon dissociations, in
   the sense of the successive absorptions of several photons before the species could completely cool down, are taken 
   into account for every dissociation process. We applied this model to four compact PAHs, namely coronene (\ch{24}{12}), 
   circumcoronene (\ch{54}{18}), circumovalene (\ch{66}{20}) and circumcircumcoronene (\ch{96}{24}), in the conditions
   of NGC 7023, both in the NW PDR and the cavity, and of the diffuse ISM.
   
   The calculated spatial evolution of the PAH charge state is in good agreement with the observational constraints. The 
   hydrogenation state is more difficult to predict as it depends on a wealth of processes, many of which have not or 
   poorly been studied in the laboratory or theoretically. Clear observational diagnostics of the hydrogenation state of 
   PAHs are also missing. Despite these difficulties, we showed that the general behaviour of PAH evolution is quite 
   robust with molecular size and with the assumptions regarding the molecular processes. It leads to four major 
   populations: superhydrogenated PAHs, normally hydrogenated PAHs, partially dehydrogenated PAHs and carbon clusters 
   (i.e. totally dehydrogenated PAHs). 
   
   Partially dehydrogenated PAHs are found to be minor species in terms of abundance regardless of the molecular size or 
   the position in the PDR. Superhydrogenated PAHs dominate only for very large PAHs ($N_{\rm C} \gtrsim 96$) in the 
   conditions of the NGC 7023 NW PDR. The size  limit below which normally hydrogenated PAHs fully dehydrogenate to give 
   birth to carbon clusters is found to lay between $N_{\rm C} \sim 50$ and 70 at the emission peak of AIBs, depending 
   on the reaction rates of PAHs with H and \hhp. In the more diffuse medium of the cavity, where the fullerene C$_{60}$ 
   was recently observed, we predict all the four species studied here to be fully dehydrogenated.
   
   We evaluated that the electronic recombination of PAH cations with electrons and the reactivity of PAHs with \hh are
   the main contributors to the uncertainties on our results. 

   In addition, we were able to rationalize the non detection of corannulene \ch{20}{10} in the Red Rectangle nebula and 
   of hexa-peri-hexabenzocoronene \ch{42}{18} in the diffuse ISM. Therefore future search for the identification of 
   individual PAHs should focus on larger species, bearing at least 50-60 carbon atoms. 
   
   More generally, multi-photon events were found to dominate the evolution of interstellar PAHs, even for relatively 
   low radiation fields as prevailing in the diffuse ISM. This results in a complex behaviour with variations of $\nH$ 
   and $G_0$, that cannot be simply parametrised with the ratio $G_0/\nH$. 
   
   We also determined PAH evolution timescales to be much longer than previously thought. In some cases, timescales 
   comparable to dynamical evolution timescales are reached, raising the question of the coupling between dynamics and 
   chemical evolution of PAHs. This could impact both the physics and chemistry of PDRs, for example through 
   photoelectric heating and \hh formation on PAHs. An other important consequence of these long timescales is the
   fact that reactions with species like O and C$^+$ may occur at similar or shorter timescales and therefore require
   more investigation.
   
   Finally, we showed that a large population of carbon clusters could exist in PDRs as a result of the dehydrogenation 
   of PAHs. Our results therefore call for fundamental studies on carbon clusters in order to determine to which extent
   these species contribute to the mid-IR emission, and affect the physical and chemical evolution of PDRs. These 
   studies are also necessary to unveil the link between PAHs and fullerenes.\\


{\it Acknowledgements.}
J. Montillaud acknowledges the support of the French Agence Nationale de la Recherche (ANR), under grant GASPARIM 
"Gas-phase PAH research for the interstellar medium", as well as the support of the Academy of Finland grant No.
250741.

This work was supported by the French National Program
Physique et Chimie du Milieu Interstellaire, which is gratefully
acknowledged.

Some of the data presented in this paper were obtained from the Multimission Archive at the Space Telescope Science 
Institute (MAST). STScI is operated by the Association of Universities for Research in Astronomy, Inc., under NASA 
contract NAS5-26555. Support for MAST for non-HST data is provided by the NASA Office of Space Science via grant 
NAG5-7584 and by other grants and contracts.

\appendix

\section{Modelling the internal energy of PAHs}\label{anx:Eint}

   In this work, we model the time evolution of both the abundances and internal energies of the studied species. The 
   latter is described as a histogram in which each bin $i$ represents the number per unit of volume of species bearing 
   an internal energy between $E_{i-1} = i \Delta E$ and $E_i = (i+1) \Delta E$, where $\Delta E$ was set to 0.25 eV. 
   In this appendix, we detail how we relate the rates characterising the fluxes of species from one bin to another,
   which are driven by UV-visible photon absorption, IR photon emission, and molecule dissociation.

   \subsection{Heating molecules}
   Species are heated after the non-ionizing absorption of a UV-visible photon. This leads to a transition from one bin 
   $i$ to a higher bin $i^{\prime}$ ($i^{\prime} > i$), at a rate which depends on the flux of UV-visible photons 
   ${\cal F}(h\nu)$, the absorption cross-section of the species $\sigma_{\rm abs}(h\nu)$, and the non-ionization yield 
   $1-Y_{\rm ion}(h\nu)$. It also depends on the triangle shaped probability function ${\cal P}_{i,i^{\prime}}(h\nu)$.
   This function expresses the probability that the energy $h\nu$ of the photon falls between the minimum ($E_{i^{\prime}-1}-E_{i}$) 
   and maximum ($E_{i^{\prime}}-E_{i-1}$) differences between the boundaries of bins $i$ and $i^{\prime}$. Therefore, 
   the rate of transitions from bin $i$ to bin $i^{\prime}$ is given by:
   \begin{equation}
      \label{eq:calc_abs_split}
      k_{\rm abs}^{i,i^{\prime}} = \int_{E_{i^{\prime}-1}-E_{i}}^{E_{i^{\prime}}-E_{i-1}} 
         {\cal P}_{i,i^{\prime}}(h\nu) \, \left(1-Y_{\rm ion}(h\nu)\right) \, \sigma_{\rm abs}(h\nu) \, {\cal F}(h\nu) \, d(h\nu)
   \end{equation}
   This expression is accurate, as long as one can neglect the variations of $\sigma_{\rm abs}$ and $Y_{\rm ion}$ with 
   the internal energy of the species, and the variations of abundance within the bin $i$. The latter condition 
   requires $\Delta E$ to be smaller than $h\nu$ for the less energetic absorbed photons that have a significant impact 
   on the internal energy of the species. This is a much weaker constraint that the conditions for a proper description 
   of the cooling of molecules.


   \subsection{Cooling molecules}
   Species cool down through the emission of IR photons. This corresponds to the transition from one bin $i^{\prime}$ to 
   a lower bin $i$ ($i^{\prime} > i$), at a rate that depends both on the IR emission rate and on the energy carried by
   the emitted photon which has to fall between the minimum and maximum differences between the boundaries of bins $i$ 
   and $i^{\prime}$. We define the factor $\delta_{i,i^{\prime}}(E)$ that equals $1$ when the latter condition is true, 
   and $0$ otherwise. For the sake of simplicity, and to ensure the computations to be tractable, we neglect the 
   dispersion of the IR energies of the emitted photons and compute $\delta_{i,i^{\prime}}(E)$ using the mean IR energy 
   as presented in Sect.~\ref{kIR}. Therefore, the effective cooling rate from bin $i^{\prime}$ to bin $i$ is given by:
   \begin{equation}
      \label{eq:calc_IR}
      \kIR^{i^{\prime},i} = \int_{E_{i^{\prime}-1}}^{E_{i^{\prime}}} \delta_{i,i^{\prime}}(E) \kIR(E) dE
   \end{equation}
   This formulation requires to use bins thinner than the less energetic IR photon to ensure that each emission leads to
   an actual cooling.

   \subsection{Molecule dissociation}
   The dissociation rate from the bin $i$ of a species is computed as the mean value of the dissociation rate of the 
   species over the width $\Delta E$ of the bin: 
   \begin{equation}
      \label{eq:kdiss_bin}
      \kdiss^i = \int_{i\Delta E}^{{(i+1)\Delta E}} \kdiss(E) ~dE / \Delta E
   \end{equation}
   This approximate expression is better for smaller energy bins.
   
   \subsection{Width of energy bins} \label{anx:binwidth}
   We empirically found that $\Delta E = 0.25$ eV provides abundances accurate at $\sim5$ \% of the asymptotic results 
   obtained when $\Delta E \rightarrow 0$. This value is therefore a good compromise between accuracy and computing time.

\bibliographystyle{aa}
\bibliography{hydroPAH}

\end{document}